\documentclass[%
 reprint,
 amsmath,amssymb,
 aps,
]{revtex4-1}

\usepackage{graphicx}
\usepackage{dcolumn}
\usepackage{bm}

\begin{document}


\title{Understanding the magnetic interactions of the zig-zag honeycomb lattice: Application to $\alpha$-RuCl$_3$}

\author{E.M. Wilson and J.T Haraldsen}
\affiliation{%
Department of Physics, University of North Florida, Jacksonville, FL 32224, USA}


\date{\today}

\begin{abstract}

This investigation covers the effects of variable exchange interactions on the spin dynamics of the zig-zag honeycomb lattice. Using a Holstein-Primakoff expansion of the Heisenberg Hamiltonian with easy-axis anisotropy, we characterize the effects of multiple nearest-neighbor and next-nearest-neighbor interactions with asymmetry within the context of a frustrated and non-frustrated zig-zag magnetic configuration. Furthermore, we compare to the known inelastic neutron scattering data for the proximate quantum spin liquid $\alpha$-RuCl$_3$, and we provide insight into the evolution of the spin dynamics, showing that the Heisenberg interaction dominates the majority of the spin excitation behavior. By analyzing the frustrated system with multiple interactions, direction-dependent Dirac nodes present themselves, and we can demonstrate that a standard Heisenberg model can accurately describe the observed magnon spectra. 

\end{abstract}

\maketitle


\section{Introduction}


The dramatic increase in the use of electronic devices in all aspects of the world has fueled a great need for high-efficiency materials. As the electronic demand increases with faster data consumption, cryptocurrency, and high-speed internet use, the energy demand and carbon footprint for electronics put them in the spotlight. In pursuing more energy-efficient electronics, the best way to improve devices is to have better materials that provide the same or better ability to store and transfer information as current electronic devices.

Areas of interest that have gained attention in the last few decades are the fields of spintronics and magnonics, which manipulate the property of spin and spin waves for information transport. Since magnetic interactions ($\sim$1-10 meV) are typically smaller than electronic excitations ($\sim$1 eV), spintronic and magnonic devices have the potential to provide considerable energy savings while maintaining or even exceeding the speed and fidelity of standard electronic components.

To identify materials that allow for the utilization of magnons and spin waves for application purposes, it is crucial to understand how magnetic interactions affect the propagation of spin excitations through various lattice configurations.

Previous studies have provided insight into square, hexagonal, honeycomb, and Kagome lattices. Each lattice configuration has various magnetic configurations consisting of ferromagnetic (FM) and antiferromagnetic (AFM) interactions. The honeycomb and Kagome lattices are particularly interesting due to the multiple sublattice interactions that can lead to Dirac nodes and potential exotic spin states like the elusive quantum spin liquid.

Since the discovery of graphene\cite{Novoselov04AAAS,Geim07NatMat}, the interest in two-dimensional systems has been promulgated due to the fascinating nature of its properties and potential applications. Like such, many new two-dimensional materials have fallen under investigation and the production of multiple honeycomb systems with similar properties have come to light such as Na$_2$IrO$_3$, RuCl$_3$, CrX$_3$ (X = Cl, Br, I), and more\cite{Baner16PRB,Chaloupka13PRL,Chaloupka16PRB,Fransson16PRB,Singh10PRB}. With the two-dimensional honeycomb lattice on the rise, we take a deeper look into the generalized honeycomb's magnetic structure and how the magnetic interactions respond to outside perturbation. 

\begin{figure}
    \centering
    \includegraphics[width=3.25in]{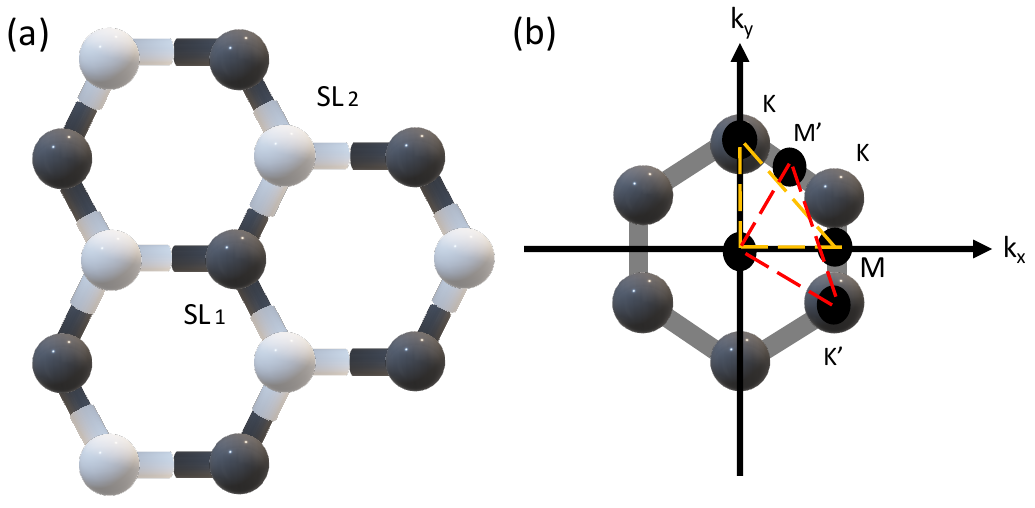}
    \caption{(a) An illustration showing the physical configuration of the honeycomb lattice as a two sublattice (2SL) hexagonal structure. (b) The reciprocal space representation of the honeycomb lattice and the high symmetry pathways $\Gamma$-M-K-$\Gamma$-M'-K'-$\Gamma$. }
    \label{fig:BZ}
\end{figure}

In this study, we focus on the zig-zag (ZZ) AFM configuration of the honeycomb lattice and use a Heisenberg spin Hamiltonian to investigate the effects of non-frustrated and frustrated interactions on the spin-wave excitations. Through a Holstein-Primakoff expansion, 


As shown in Fig. \ref{fig:BZ}(a), the honeycomb lattice is a unique form of hexagonal lattice that consists of two sublattices (2SL), where the primitive translation vectors of the hexagonal lattice introduce angles of 120$^{\cdot}$ in between equal lengths and rotate 90$^{\cdot}$ with length $b=4\pi/a\sqrt{3}$ within reciprocal space. Such a structure presents five different collinear (and other non-collinear) magnetic structures such as ferromagnetic (FM), antiferromagnetic (AFM), zigzag (ZZ), dimerized (DIM), and armchair (ARM)\cite{Boyko18PRB,cayssol13CRP}; however, the focus of this study is on the zig-zag magnetic configuration.

To begin understanding the zig-zag magnetic configuration and how it behaves to disturbance, it is necessary to note the behavior shown in the most basic FM and AFM configurations. It is observed that the introduction of spin into the honeycomb lattice complicates the presence of inversion symmetry in the two sub-lattice structures. Observing the FM configuration, it is seen that in every direction, inversion symmetry is maintained and, therefore, should produce a Dirac cone\cite{Gao21PRB}. Interestingly, in the AFM configuration, no matter what direction is analyzed, when the magnetic structure is imposed, it is found that inversion symmetry is broken and produces a 180-degree rotation. This rotation forces $SL_1=SL_2$ and produces only a single-band mode within its spin-wave spectra and has been discussed further in ref. [\cite{Boyko18PRB}-\cite{cayssol13CRP},\cite{Cheng16PRL}-\cite{Fransson16PRB})]. If the FM configuration maintains inversion symmetry while the AFM configuration breaks it, what would a system with these two alternating interactions look like?

\section{Zig-Zag Spin Hamiltonian}

 Understanding the zigzag (ZZ) collinear magnetic structure present within a honeycomb lattice starts with a generalized Heisenberg model. While invoking spin on the magnetic configuration, nearest and next nearest neighbor interactions are modeled to reproduce magnon excitation. The Heisenberg Hamiltonian capturing these interactions is expressed as:

\begin{equation}
    H=-\frac{1}{2}\sum_{i\neq j}\alpha_{ij}\beta_{ij} J_{ij}\vec{S_{i}}\cdot\vec{S_{j}}-D\sum_{i}S^2_{iz}
\end{equation}

Where \(J_{ij}\) signifies the exchange interactions between spins \(S_{i}\) and \(S_{j}\) at corresponding sites \(i\) and \(j\). The term \(D\) accounts for easy-axis anisotropy. The coefficients \(\alpha\) and \(\beta\) are scaling factors for the nearest and next-nearest neighbor exchange interactions. Positive scaling factors correspond to ferromagnetic interactions, while negative ones indicate antiferromagnetic interactions. Moreover, the prime values of these interactions, \(\alpha'\) and \(\beta'\), introduce asymmetry to model imperfections, which plays a pivotal role in describing the zigzag phase due to the potential competition between \(J_1\) and \(J_2\) \cite{Haraldsen09JoP}. This complexity leads to inversion symmetry being retained solely between two nearest neighbor spins, yielding a total of only two modes\cite{Boyko18PRB}.

\begin{figure}
    \centering
    \includegraphics[width=8cm]{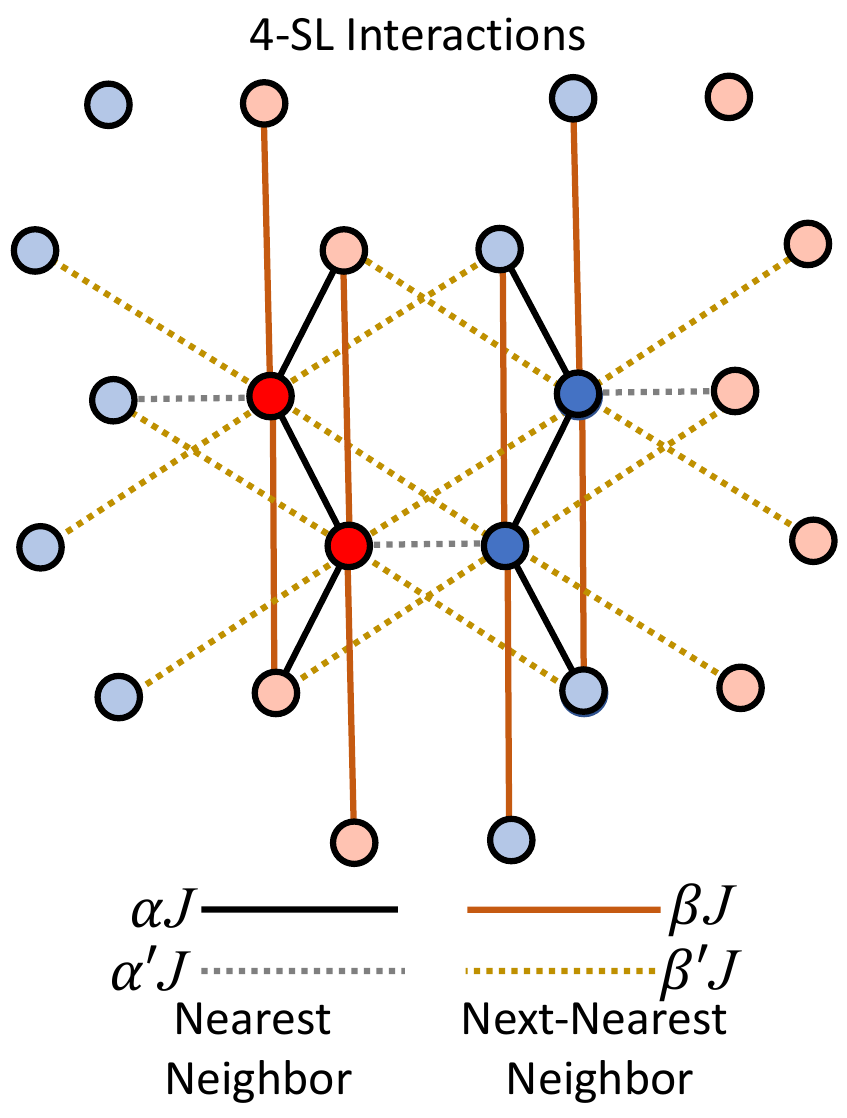}
    \caption{ Four sub-lattice interactions where $\alpha$ is the nearest neighbor interaction and $\beta$ is the next nearest neighbor interaction. The prime terms $\alpha$' and $\beta$' are asymmetric nearest and next nearest neighbor interactions respectively. }
    \label{fig:lattice}
\end{figure}

The zigzag magnetic configuration, characterized by a collinear arrangement of spins, demonstrates a unique pattern where nearest and next-nearest neighboring atoms interact along a distinctive 'zig-zag' path (see Fig. \ref{fig:lattice}). This striking configuration dynamically oscillates between ferromagnetic (FM) and antiferromagnetic (AFM) interactions. Intriguingly, within the framework of the zigzag magnetic arrangement, a noteworthy phenomenon emerges: when observed in a specific direction, inversion symmetry is perturbed, echoing the behavior seen in AFM configurations. Moreover, inversion symmetry remains preserved along two distinct directions upon system rotation. This observation holds the potential to uncover the existence of magnetic Dirac nodes, arising from the intricate interplay of direction-dependent inversion symmetry breaking \cite{Haraldsen09PRL}\cite{Haraldsen09JoP}\cite{Toth15JoP}\cite{AuerbachAPS2018}. The effective Hamiltonian for this model, then is

\begin{equation}
\begin{split}
H = -\frac{1}{2}\sum_{\langle i,j \rangle} \alpha J_1\vec{S_{i}}\cdot\vec{S_{j}} 
-\frac{1}{2}\sum_{\langle\langle i,j \rangle\rangle} \beta J_2\vec{S_{i}}\cdot\vec{S_{j}} \\
-\frac{1}{2}\sum_{\langle i,j \rangle} \alpha' J_1\vec{S_{i}}\cdot\vec{S_{j}} 
-\frac{1}{2}\sum_{\langle\langle i,j \rangle\rangle} \beta' J_2\vec{S_{i}}\cdot\vec{S_{j}} 
- D\sum_{i}S^2_{iz}
\end{split}
\end{equation}

The terms involving \(\alpha\) and \(\beta\) correspond to the nearest and next nearest neighbor interactions, respectively, with corresponding interaction strengths \(J_1\) and \(J_2\). The prime values \(\alpha'\) and \(\beta'\) are introduced to account for imperfections in these interactions. The final term represents the single-ion anisotropy, \(D\), associated with each spin site \(i\). This comprehensive Hamiltonian captures the intricate interplay between exchange interactions and anisotropy, crucial for understanding the magnetic properties of the system; however, an expansion of the terms is still necessary for capturing the dispersion relation of spin wave excitation within a bosonic system\cite{Cabra11PRB}. 

To accomplish this, the Holstein-Primakoff (HP) expansion is employed on the Heisenberg spin Hamiltonian, allowing us to delve into the effects of nearest-neighbor and next-nearest-neighbor interactions\cite{AuerbachAPS2018}. The HP transformation enables us to express the spin operators as combinations of bosonic creation (\(b_i^\dagger\)) and annihilation (\(b_i\)) operators, which correspond to magnons, the quanta of spin excitations. The HP transformation involves mapping the spin operators onto these bosonic operators and then truncating the expansion at certain orders in \(1/S\)\cite{Holstein40PR}\cite{Cabra11PRB}\cite{Fouet01EPJB}. 
\begin{equation}
    S_i^{\pm} = S_i^{x}\pm iS_i^{y}
\end{equation}
where the quantum spin operators \(S_{ix}\) and \(S_{iy}\) transform into a combination of bosonic creation and annihilation operators. The \(S_{i}^{+}\) operator represents the raising operator, and \(S_{i}^{-}\) represents the lowering operator. Together, they create a coherent superposition of spin states. This transformation allows us to describe spin excitations (magnons) as bosonic quanta\cite{Nieto97PoP}.

\begin{equation}
    S_i^{z} = S_i - b_i^{\dagger}b_i 
\end{equation}
\(S_i^{z}\) represents the \(z\)-component of the spin operator \(S_i\), which corresponds to the quantum spin at site \(i\). The right-hand side of the equation reveals that we can represent the \(z\)-component operator by using the spin operator \(S_i\) itself along with the number operator \(b_i^{\dagger}b_i\). This term involving the number operator signifies the magnon number operator, responsible for counting the number of magnons present at site \(i\) \cite{Imamura04PRB}.

\begin{equation}
    S_i^{+} = (\sqrt{2S_i-b_i^{\dagger}}b_i)b_i\approx \sqrt{2S_i} b_i
\end{equation}

 We express \(S_i^{+}\) by employing the creation and annihilation operators \(b_i^{\dagger}\) and \(b_i\). Notably, the equation incorporates the square root of \(2S_i - b_i^{\dagger}\), originating from the fact that magnons emerge as a result of spins transitioning from a lower-energy state to a higher-energy state\cite{Fouet01EPJB}.

\begin{equation}
    S_i^{-} = b_i^{\dagger}\sqrt{2S_i-b_i^{\dagger}}b_i\approx \sqrt{2S_i} b_i^{\dagger}
\end{equation}
Similarly, this equation addresses the \(S_i^{-}\) operator, responsible for lowering the spin state at site \(i\) by one unit. The equation is the conjugate of the previous equation and encompasses the use of the creation and annihilation operators \(b_i^{\dagger}\) and \(b_i\). Both simplifications come from expanding the square root term using a Taylor series up to the first-order term. When \(S\) is large, quantum fluctuations become less significant, and classical-like behaviors emerge. Leveraging this characteristic, the \(1/S\) expansion capitalizes on the largeness of \(S\) to treat \(1/S\) as a small parameter, enabling the development of a systematic perturbed expansion\cite{Cabra11PRB}\cite{Holstein40PR}. This expansion takes the form of a series: 
\begin{equation}
    H = E_0+H_1+H_2+..., 
\end{equation}
where $E_0$ is the classical energy, $H_1$ is the vacuum energy, and $H_2$ is the resultant spin-dynamics of the system.
A classical energy state emerges for our specific configuration within the framework of this \(1/S\) expansion. Specifically, the expression for the classical energy is given by:

\begin{equation}
    E_{zz}=-\frac{S^2}{2}(J_1 -2J_2)
\end{equation}
Intriguingly, including higher-order exchange parameters introduces competition within the zigzag (ZZ) configuration. Consequently, the stability of this configuration is not guaranteed without an easy-axis anisotropy \cite{Haraldsen09JoP}\cite{Trumper00PRB}. Notably, despite these complexities, vacuum fluctuations and temperature continue to wield influence over the boundaries and characteristics associated with the zigzag magnetic phase, as explored in the study by Fransson et al. \cite{Fransson16PRB}.

\section{\label{sec:level2} Effects of Nearest and Next Nearest Neighbors} 

\begin{figure}[htp]
    \centering
    \includegraphics[width= 8.5 cm]{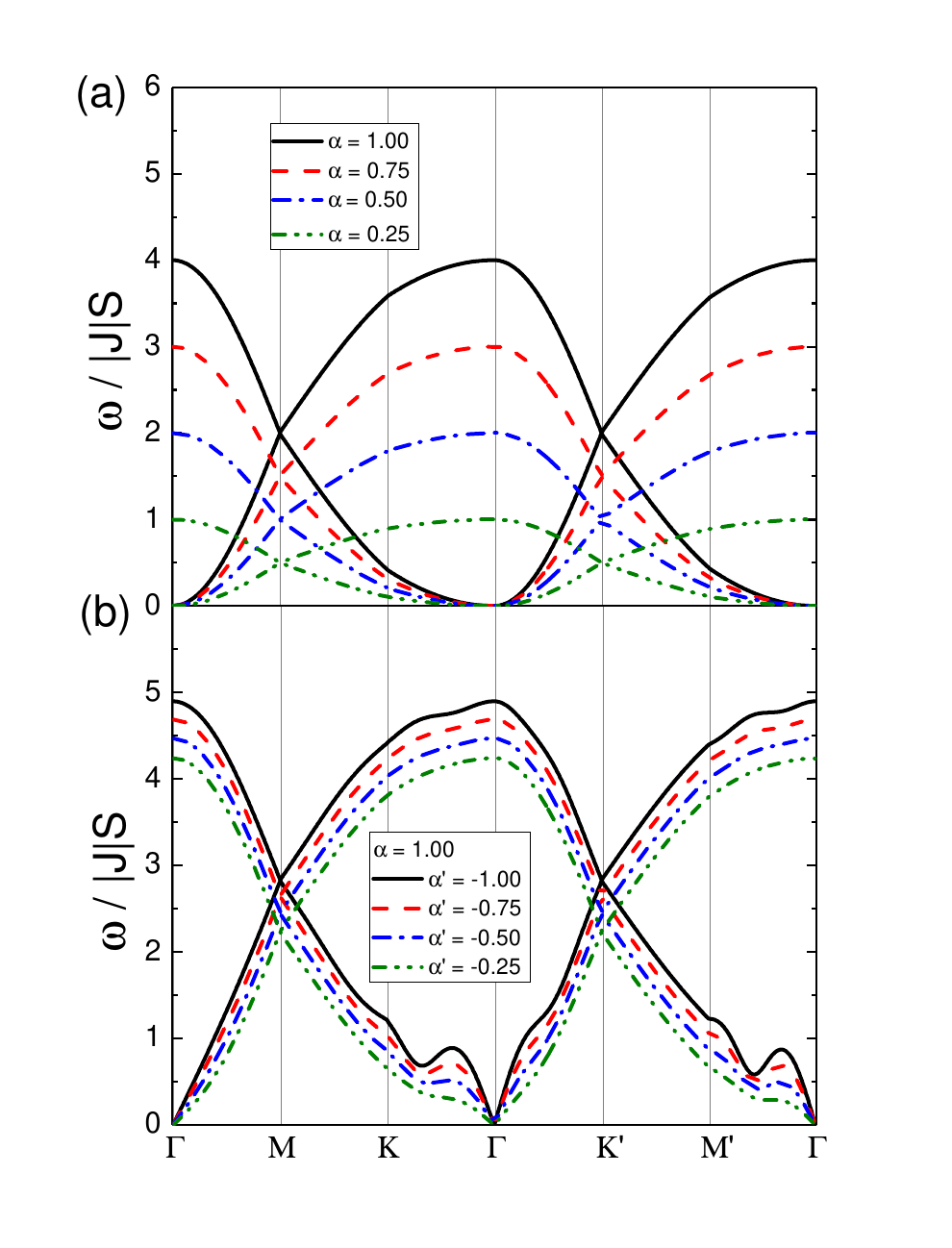}
    \caption{2D plot showing the nearest neighbor interaction's behavior is shown in the top (a) as $\alpha$ is added. This produces two direction-dependent diagonal nodes at M and K'. The bottom plot (b) shows how this changes as deformities are introduced as $\alpha'$. This creates distortions in the spin-wave that increase alongside $\alpha$'.}
    \label{fig:Nearest Neighbor interaction}
\end{figure}

Utilizing known regions of stability for magnetic phases, diagonalization of a dynamics matrix composed of superexchange parameters arising from spin-orbit coupling produces the observed mixing of magnon modes to be analyzed\cite{Colpa78Physica}\cite{Ozel19PRB}\cite{Nieto97PoP}. To show the evolution of the spin dynamics represented by a ZZ magnetic configuration within a honeycomb lattice, we start with a base case where the nearest neighbor interaction $\alpha$ $>$ 0, and all other interactions are set to 0. In this regime, the ferromagnetic configuration dominates. The zigzag configuration consists of alternating stripes of up and down spins along the zigzag direction of the honeycomb lattice. This alters spin exchanges along the zigzag direction, which breaks inversion symmetry and produces a $180^o$ rotation\cite{Boyko18PRB}. Given such, the behavior of the ferromagnetic interaction $\alpha$ along the ZZ is produced with stability, as seen in Fig. \ref{fig:Nearest Neighbor interaction}. As $\alpha$ increases, the energy dispersion range increases along the high-symmetry pathway $\Gamma$ to M to K and back to $\Gamma$. The resultant spin-wave progresses from a smooth behavior to a more deformed propagation, most noticeably from M to K' as the magnitude of $\alpha$ increases. Inversion symmetry exists only among two nearest neighbor spins within this ZZ magnetic configuration. Observably, this produces a resultant spin wave with only two observed modes. Distinct crossing points between the two spin-wave modes arise at high-symmetry points M and K'. The aforementioned implicates magnetic Dirac nodes at these points as is discussed in Ref[\cite{Fransson16PRB} \cite{Gao21PRB}]. 

Furthering our investigation of the ZZ magnetic configuration, the asymmetry within the nearest neighboring interaction $\alpha$ interaction is introduced through $\alpha$'. As $\alpha =1.00$ remains ferromagnetic, the natural order is to introduce $\alpha'$ to be anti-ferromagnetic. This provides insight into the role $\alpha'$ plays within the model and how it affects the behavior reflected in the first case where there was no $\alpha'$. 

Starting with $\lvert\alpha\rvert$$>$$\lvert\alpha'\rvert$, small values of $\alpha$' are added and steadily increased until $\alpha$'$=$-$\alpha$. As the values increase, the noted deformities in Fig.\ref{fig:Nearest Neighbor interaction}(a) become more profound, and new distinct behavior blatantly presents itself from K to $\Gamma'$ and M' to $\Gamma$. It is observed that the stability increases with the introduction of $\alpha$' as the high-symmetry points become more recognizable. Moreover, the spin-wave produced as it propagates throughout the Brillouin zone becomes more defined as its energy range increases. The spin waves produced within the Brillouin zone have higher dispersion energy associated with higher asymmetry, and it's observed as $\alpha$' increases, so does the dispersion energy. This dispersion along the high symmetry pathway retained the magnetic Dirac nodes at M and K'. Building the model by its components gives insight into how each interaction affects the total behavior and their role in the spin dynamics of a ZZ magnetic structure. 

\begin{figure}[htp]
    \centering
    \includegraphics[width=8.5 cm]{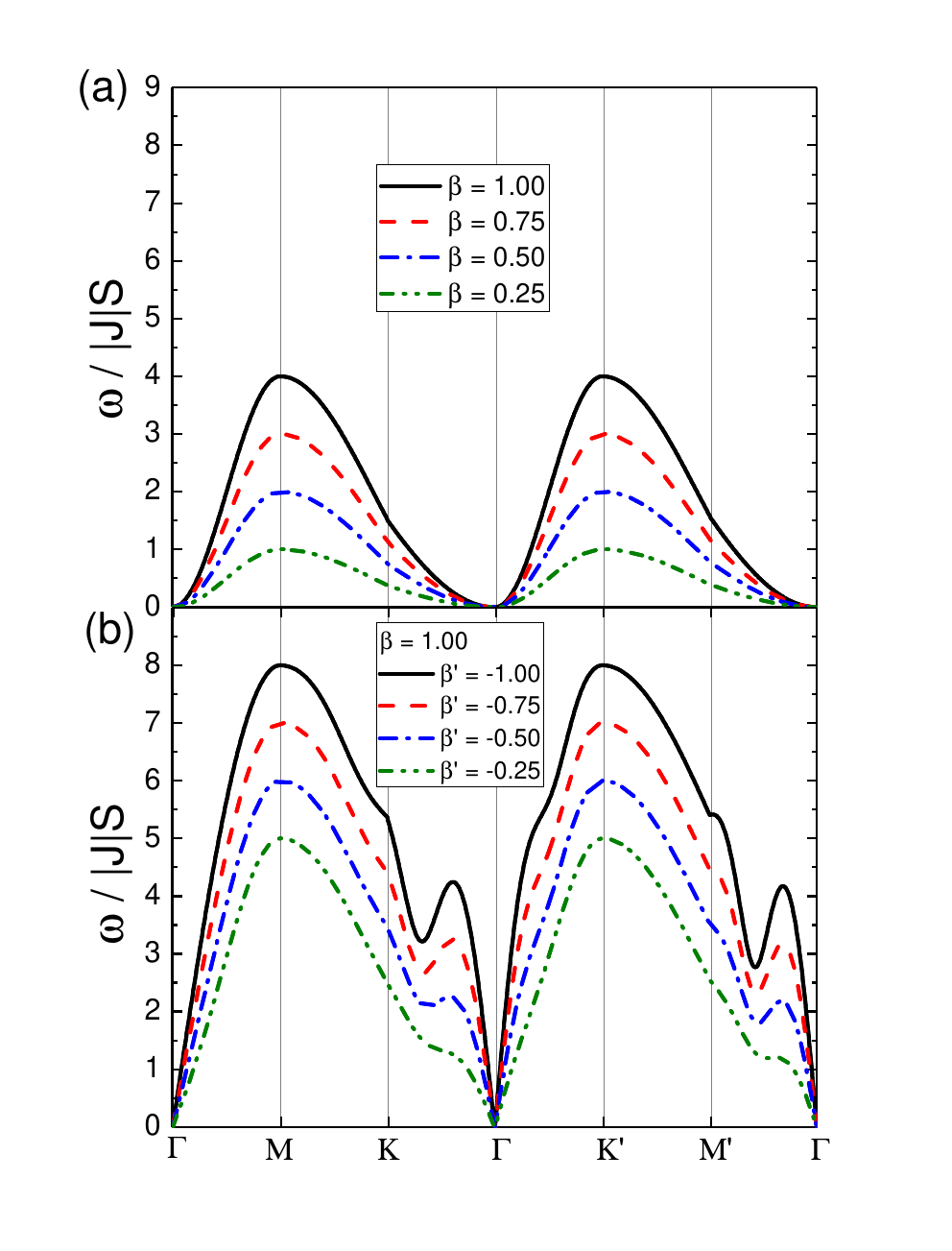}
    \caption{2D plot showing the next nearest neighbor interaction's behavior is shown in the top (a) as $\beta$ is added. The bottom plot (b), shows how this changes as deformities are introduced as $\beta'$. This creates distortions in the spin wave and increases alongside $\beta '$. Only one spin wave is produced as the interactions of the next nearest neighbor are degenerate in this direction.}
    \label{fig:Next Nearest Neighbor interaction}
\end{figure}

Continuing with the same analytic method, the next nearest neighbor interaction is examined by itself with $\beta > 0$, and all other interactions are set to 0. Noticeably, the resultant spin-wave has only one mode; therefore, there is no possibility of crossover points producing Dirac nodes. The emergence of singular mode is a product of spins flipping between sub-lattices and giving rise to multiple modes cohabiting with the same energy levels. The modes mentioned above show great stability, and as $\beta$ increases, the energy increases, producing more distinct peaks at M and K'. Switching to $\beta < 0$, the same dispersion was observed with $\beta > 0$. The change for the next nearest-neighbor interaction is reflected in Fig.\ref{fig:Next Nearest Neighbor interaction}(a) The dispersion invoked by the next nearest neighbor interaction is double degenerate and would explain the interaction not changing for its spin.

Similarly to the case where the nearest neighbor interaction was examined, asymmetry is added to the system by introducing $\beta'$. Setting $\beta > 0$ increases asymmetry and starts to add distinct behavior to the interaction. As the dispersion approaches $\Gamma$ from both K and M', a spike is introduced before dropping into $\Gamma$, as observed in Fig.\ref{fig:Next Nearest Neighbor interaction}(b). This shows that no Dirac modes are produced through this interaction, similar to AFM configurations such as in ref[\cite{Cheng16PRL}]. Seemingly, the next nearest neighbor interaction has strong effects to systems such as the AFM configuration and, therefore, could play a dominant role in the ZZ magnetic configuration due to the double degeneracy. Determining the evolution of a zigzag magnetic configuration's spin dynamics in a honeycomb lattice depends on all the constituent parts. By analyzing these parts individually, insight into the interactions and how they affect the resultant perturbation is prevalent.  

\subsection{Non-Frustrated Model}

\begin{figure*}
    \centering
    \includegraphics[width=7.0in]{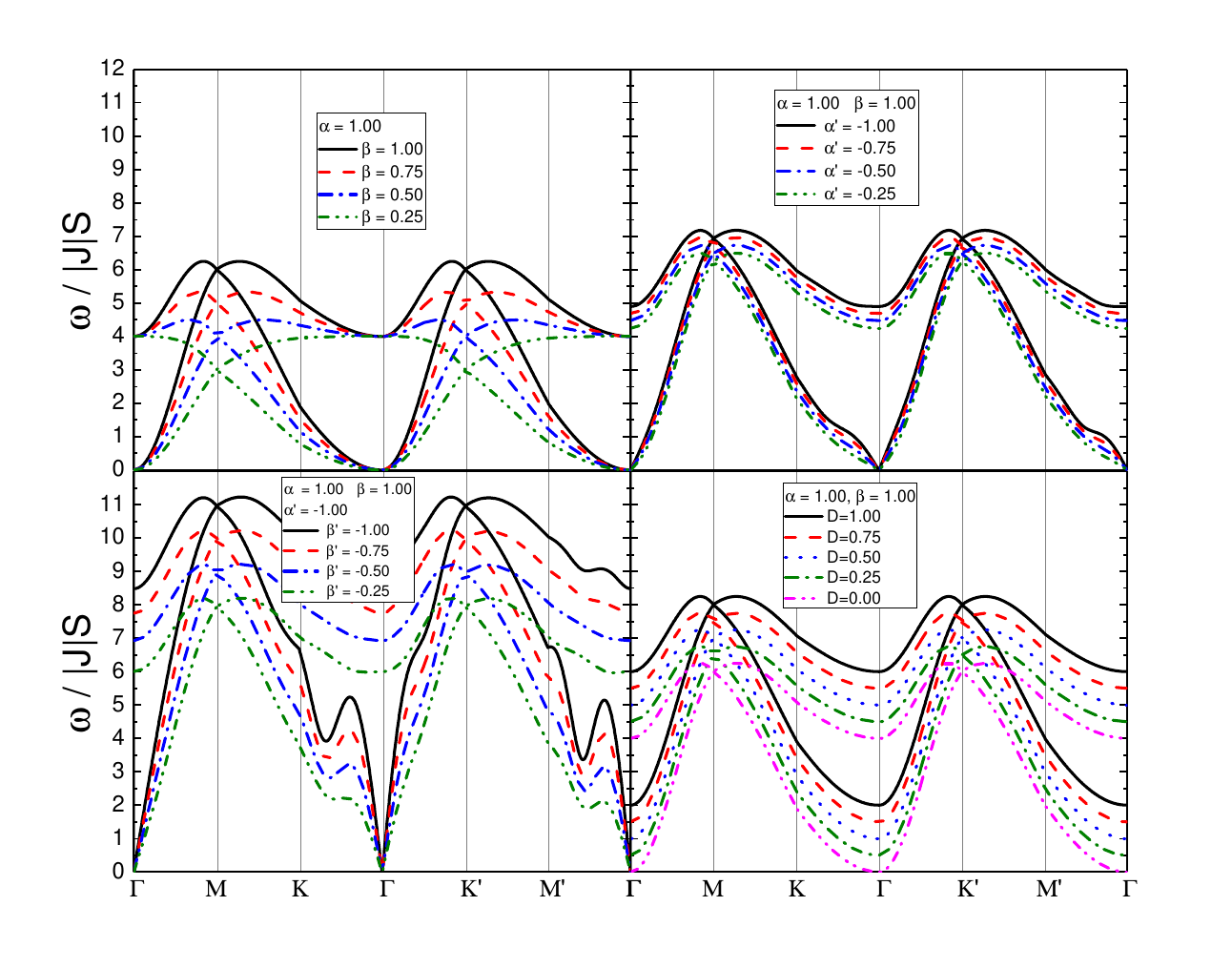}
    \caption{2D plot of the evolution of spin waves within the non-frustrated model as the interactions change. Each plot starts dotted, and as more data is added, it becomes more solid until the final calculation, which is the solid black. Starting in the top left with $\alpha$=-1.00, $\beta$ is added in increments of -0.25 until it reaches $\beta$=-1.00. In the top right, $\alpha '$ produces deformities and is added similarly. This adds small changes in the behavior but not much energy. As $\beta '$ is introduced in the bottom left, the changes in the spin-wave become more accented, and the deformities more prevalent with higher energies. An anisotropy test in the bottom right shows how it affects the calculations. As anisotropy is added, the energy of the spin-wave increases all around without changing the behavior. }
    \label{fig:Non-frustrated-Model}
\end{figure*}

Within the non-frustrated model the interactions coalesce without competition where they are "content" in their configuration. Starting at the nearest neighbor interaction $\alpha$, a next nearest neighbor interaction$\beta$ is added. Observing how the response to disturbance changes with the addition of another interaction, the nearest neighbor interaction is set to a FM interaction $\alpha$= 1.00. The FM next nearest neighbor interaction $\beta$ is then added in increments of 0.25 until $\alpha$=$\beta$. Looking at Fig.\ref{fig:Non-frustrated-Model}, the top left plot illustrates the calculations as the interactions approach being equal, starting as a dashed line and progressing to a solid line with the increase of $\beta$. What is observed is the overall energy spectrum increase at the peaks M and K' as well as the troughs at $\Gamma$ drop in energy and become more defined. This suggests so far that between the two, the nearest neighbor interaction dictates the behavior observed, and the addition of the next nearest neighbor increases the strength of the spin-wave and controls the presented crossover points. This is potentially due to the degeneracy presented in Fig.\ref{fig:Next Nearest Neighbor interaction}. where all the spin waves have the same energy levels, and only one perturbation is preserved. The produced spin wave by $\alpha$ is then shaped by the degeneracy introduced by $\beta$. The crossover points are presented on high symmetry locations M and K' at ~6 meV as $\alpha$ = $\beta$.

The next panel on the top right of Fig.\ref{fig:Non-frustrated-Model} continues from the solid black line in the top left panel. With $\alpha$=$\beta$, a second nearest neighbor interaction called $\alpha$' is added and is set to be AFM. As $\alpha$' increases by increments of -0.25 until $\alpha$'= -1.00, the general characteristics of the spin-wave don't change. Instead, what is observed is a small gradual increase in the asymmetry as the behavior itself becomes more defined and the overall energy increases. Moreover, deformities begin to present themselves from K-$\Gamma$ and M'-$\Gamma$, but only slightly. These deformities are expected from what was observed when looking at the introduction of $\alpha$' to $\alpha$ in Fig.\ref{fig:Nearest Neighbor interaction}. 
Furthermore, in the bottom left panel of Fig. \ref{fig:Non-frustrated-Model} the non-frustrated model is completed by the addition of asymmetry among the next nearest neighbor interactions with $\beta$'. The calculations are continued by starting with $\alpha$ = $\beta$ = - $\alpha$' and adding $\beta$' in increments of -0.25. The deformations presented from K to $\Gamma$ and M to $\Gamma$ by adding $\alpha$' become very distinct as the energy increases across the spin-wave. The non-frustrated model doesn't require easy-axis anisotropy as all the nearest and next-nearest neighbors match their configuration within the honeycomb lattice. With all interactions $\alpha$ = $\beta$ = -$\alpha$' = - $\beta$' represented in the solid black line in the bottom left panel, the crossover points are observed to shift up to ~ 11 meV, and the system continues to be stable. 

On the bottom right panel in Fig.\ref{fig:Non-frustrated-Model}, anisotropy is added similarly to the other calculations by adding increments of 0.25 to the initial case of $\alpha$ = $\beta$ = 1.00. As the single-ion anisotropy increases, the energy is proportionally increased across the spin wave without changing the observed propagation. Seemingly, the system's response to an outside disturbance within a non-frustrated model is stable without requiring any anisotropy. However, what happens as frustration is introduced to the system? If the interactions are changed to contradict their configurations within the honeycomb lattice, can the system be stable?

\subsection{Frustrated Model}
\begin{figure*}
    \centering
    \includegraphics[width=7.0in]{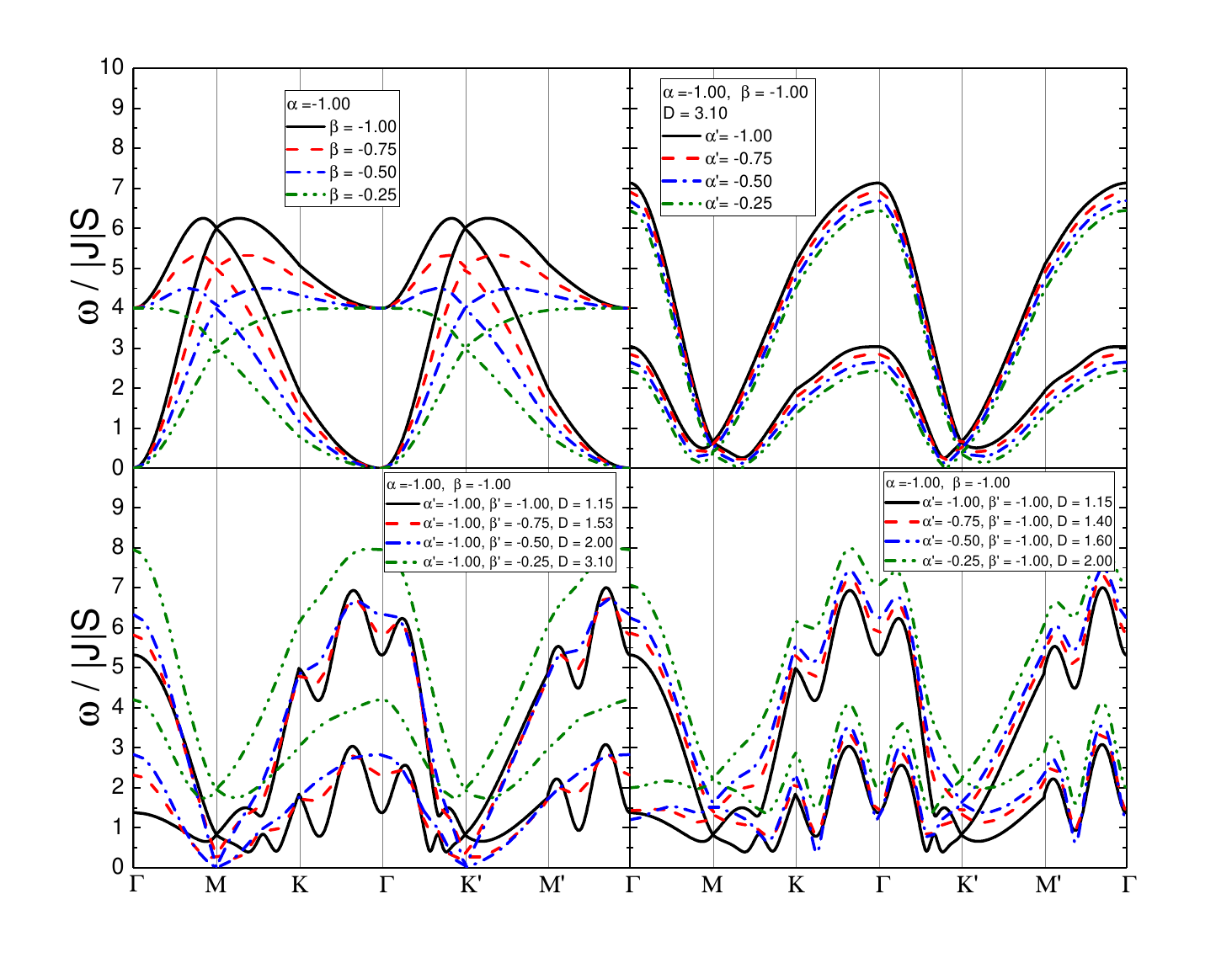}
    \caption{2D plot of the evolution of spin waves within the frustrated model as the interactions change. Each plot starts dotted, and as more data is added, it becomes more solid until the final calculation, which is the solid black.  Starting in the top left with the non-frustrated model  $\alpha$ is set to -1.00, and  $\beta$ is added in increments of -0.25 until it reaches $\beta$=-1.00. After this, we add frustration in our model through competition of exchanges through $\alpha '$. This is done in similar increments and requires anisotropy to stabilize the behavior. This also causes the Dirac nodes to drop in energy, and the overall behavior of the spin wave inverts. $\beta '$ is introduced, increased in the bottom left, and adjusted in different ratios in the bottom right. As $\beta '$ is added, the anisotropy to stabilize the system is reduced and defines the deformities that appear in the spin wave.}
    \label{fig:Frus-Model}
\end{figure*}

The honeycomb lattice, characterized by a coordination number of three, represents a unique lattice structure with bipartite properties. This means that its sites can be partitioned into two sublattices, where each site within one sublattice exclusively connects to sites in the other sublattice\cite{Tierno16PRL}. Geometric frustration emerges as the lattice's triangular arrangement conflicts with the magnetic interactions among its components. This results in frustration, where the concurrent minimization of all interactions becomes unattainable, \cite{Moessner01CJoP}\cite{Swanson09PRB}. This phenomenon is frequently encountered in systems featuring specific lattice symmetries, such as triangular or honeycomb lattices, and it gives rise to exotic phenomena like spin ice and spin liquid phases.\cite{Cabra11PRB}\cite{Beach09PRB}\cite{Haraldsen09PRL}.

A honeycomb system with identical magnetic interactions does not match the physical configuration and forces a competition of exchange interactions. As the interactions vie to be stable within their configuration, frustration is produced within the system, causing a different dispersion. The top left of Figure 6 shows the non-frustrated model where $\alpha$ and $\beta$ are AFM and $\beta$ is introduced in increments of -0.25. When $\alpha$ = $\beta$, a different AFM nearest neighbor interaction is introduced as $\alpha$', breaking the symmetry within the nearest neighbor interactions, which causes the dispersion to become completely unstable until easy axis anisotropy (D) is instituted and stabilizes the interactions at D = 3.10. The introduction of asymmetry produces an inversion of the behavior presented in the non-frustrated model, dropping the Dirac crossover point of the spin-wave from ~6 meV to ~0.8 meV as observed in the top right of Fig.\ref{fig:Frus-Model} As $\alpha$' is increased by -0.25, the behavior remains the same and the energy makes a small shift from ~6.5 meV up through to ~7.2 meV but the Dirac node remains the same. 

Continuing the trend of adding asymmetry, next-nearest neighbor asymmetry is initiated as $\beta$'. This causes the energy to shift up to ~8 meV, and as $\beta$' increases, the required anisotropy to stabilize the system decreases. Moreover, by increasing the next nearest neighbor asymmetry, the peaks of the dispersion begin to collapse, creating smaller peaks between M-K-$\Gamma$-K' and from M'-$\Gamma$ as can be seen in the bottom left of Fig.\ref{fig:Frus-Model}. To further understand the dynamics between the frustration in the interactions, a variation of how the asymmetry is introduced is established by beginning with next-nearest neighbor asymmetry and bring in nearest neighbor asymmetry in increments of -0.25. The deformation of the spin waves is initially present, and as $\alpha$' increases, the anisotropy decreases and the peaks between M-K-$\Gamma$-K' and from M'-$\Gamma$ become more defined on the bottom right of Fig.\ref{fig:Frus-Model}.  


\section{\label{sec:level2} Frustrated Model Applied  to $\alpha$-RuCl$_3$} 
\begin{figure}
    \centering
    \includegraphics[width= 10cm]{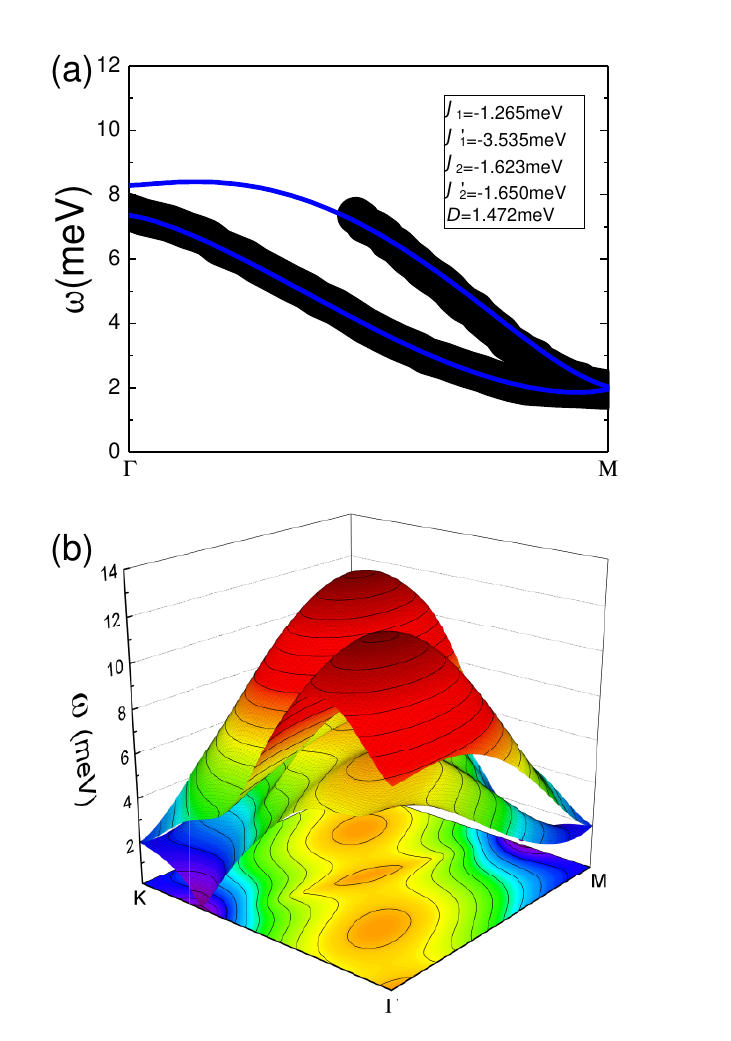}
    \caption{On the top figure (a), a 2D plot comparing an approximate experimental inelastic neutron scattering data from \cite{Ran17PRL} as the black line and the calculated spin-wave from the frustrated model as the blue line, which is shown through the high symmetry pathway $\Gamma$ - M. On the bottom figure (b), a 3D plot comparing the same data shows the spin wave throughout the entire Brillioun zone.  }
    \label{fig:RuCl3}
\end{figure}

Building upon the insights gained from our investigation into frustrated magnetic systems on a zig-zag honeycomb lattice, we apply our easy axis frustrated model to a current material of interest in the condensed matter community\cite{Johnson15PRB}\cite{Ran17PRL}. $\alpha$-RuCl$_3$ is composed of thinly interconnected layers of RuCl$_6$ octahedra that share edges, with the central Ru$^{3+}$ ions with 4d$^5$ orbital arranged in an almost perfect honeycomb configuration, a critical factor utilized in the Kitaev-Heisenberg model\cite{Roslova19InoChem}. This indicates $\alpha$-RuCl$_3$ as a prime candidate for the realization of fractionalized Kitaev physics, as well as quantum spin liquid behavior\cite{Baner16NatMat}\cite{Savary17RoPiP}\cite{Kim15PRB}\cite{Sandilands16PRB}.

Currently, there is no complete consensus on the local space group associated with $\alpha$-RuCl$_3$. A monoclinic space group C2/m at high temperatures is generally accepted as reported by Johnson et al. \cite{Johnson15PRB}; however, there have been many supportive as well as contradicting reports of different structures such as the trigonal $P3_{1}12$\cite{Fletcher67JCSA}\cite{Groenke2020BTU}, and rhombohedral R$\bar{3}$\cite{park2016emergence}\cite{Pai2021JOPCC} space groups. Seemingly sample dependent, it has recently been shown that $\alpha$-RuCl$_3$ can potentially manifest in all three polytypes and that there is a temperature dependence on the structural transitions\cite{HedaZhang2023PRM} caused by stacking faults \cite{JinZhang2023NL}. This has created a disparity in the condensed matter community because the electronic and magnetic configurations cannot be accurately resolved without an established structural space group.

Considering that $Ru^{3+}$ ions carry a spin of S=1/2, a monoclinic (C2/m) perspective allows for single ion anisotropic exchange variation between the Ru-Ru bonds in a single crystal. The magnetic moments of the ruthenium ($Ru$) ions align anti-parallel to each other along one direction and then reverse their alignment along the perpendicular direction. This results in a zigzag pattern of magnetic moments in the material\cite{Johnson15PRB}\cite{Bera17PRB}\cite{Cabra11PRB}\cite{Chaloupka16PRB}.
Given the honeycomb's propensity to induce frustration in magnetic configurations, the interactions become highly anisotropic and depend on the spins relative to the bonds between $Ru^{3+}$ ions\cite{Johnson15PRB}\cite{Tierno16PRL}.

To understand $\alpha$-RuCl$_3$'s spin-orbit excitation spectrum, various experimental techniques like THz spectroscopy, IR spectroscopy, and Raman spectroscopy have been used\cite{Loidl21JoP}. Literature investigated this spectrum, identifying the first absorption band as the spin-orbit exciton\cite{War20PRR}. Controversies exist, with some suggesting that the featureless spectra could result from magnetic anharmonicity and a breakdown of magnon excitations \cite{Winter17JoP}\cite{Winter17NatComm}\cite{Kubota15PRB}\cite{Bachus20PRL}. Nonetheless, $\alpha$-RuCl$_3$ has been demonstrated that any supplementary crystal field effects, such as trigonal or tetragonal distortions, exert negligible influence when contrasted with the prevailing octahedral crystal field\cite{Sears15PRB}. Consequently, despite its comparatively reduced bare spin-orbit coupling (SOC) value, single crystal $\alpha$-RuCl$_3$ continues to manifest significant SOC-driven effects with an ordered moment characterized by in-plane and out-of-plane anisotropy components\cite{Luo22PRB}\cite{Kim15PRB}\cite{Sandilands16PRB}\cite{Plumb14PRB}. Previous investigations suggest that in-plane anisotropy dominates the bulk crystal of $\alpha$-RuCl$_3$ and easy-axis anisotropy dominates single-layer samples\cite{Yang22NatMat}. Moreover, it has been shown that the in-plane lattice constant expands 3\% from 5.99 \AA (bulk) to 6.19 \AA (monolayer) \cite{Wang22DirectOO}. Therefore, as layers are removed from a multi-layer sample down to a single layer, $\alpha$-RuCl$_3$ transitions from easy plane to easy axis anisotropy\cite{Yang22NatMat}\cite{Toth15JoP}. 

Applying a frustrated model with asymmetric superexchange interactions and easy axis anisotropy, we analyze inelastic neutron scattering data for $\alpha$-RuCl$_3$ reported in Ref[\cite{Ran17PRL}] as shown in Fig.\ref{fig:RuCl3}. While our model does not incorporate Kitaev interactions, and the quasi-quantum 1/S expansion is limited in its ability to capture S = 1/2 characteristics fully, we employed a photometric analysis program to approximate the inelastic neutron scattering data. The primary objective was to investigate whether adjustments to model parameters could result in a close fit, as illustrated in Fig.\ref{fig:RuCl3}(a). In Fig.\ref{fig:RuCl3}, the black line corresponds to the approximated photometric data, while the general frustration model generates the blue curve. Fig.\ref{fig:RuCl3}(b) presents a 3D dispersion spanning the Brillouin zone. The breaking of inversion symmetry produces Dirac cones that are unmistakably visible as central, sharply defined conical structures. This observation confirms the presence of Dirac nodes in the material's electronic structure; moreover, in the context of the applied frustrated model for the zigzag magnetic configuration, we observe that the required breaking of inversion symmetry is produced in two separate directions for $\alpha$-RuCl$_3$. 



\section{\label{sec:level7} Conclusions }

The increasing need for advanced materials in electronic device development has spurred investigations into spintronics and magnonics. These avenues aim to conserve energy while potentially outperforming traditional electronics, resulting in numerous studies \cite{Baner16PRB}\cite{Baner16NatMat}\cite{Bera17PRB}-\cite{Fransson16PRB}\cite{Geim07NatMat}\cite{Goerbig14dirac}\cite{Kim20PRB}. Consequently, comprehending the impact of magnetic interactions on the propagation of spin excitations across diverse lattice configurations becomes paramount for identifying materials amenable to magnon and spin wave-based applications. This investigation aims to elucidate the influence of various magnetic exchange interactions on the spin dynamics inherent to the zig-zag honeycomb lattice while also emphasizing its relevance to $\alpha$-RuCl$_3$.

Particularly focusing on the zigzag configuration characterized by alternating up and down spins along its direction. The zigzag lattice arrangement disrupts spin exchanges, breaking inversion symmetry and resulting in a pronounced $180^o$ rotation. Our investigation demonstrates the stability of the ferromagnetic interaction $\alpha$ within this zigzag pattern, as illustrated in Fig. \ref{fig:Nearest Neighbor interaction}. Increasing $\alpha$ corresponds to an expanded energy dispersion range, notably observed along the high-symmetry path $\Gamma$ to M to K and back to $\Gamma$. Additionally, inversion symmetry persists only among the two nearest neighbor spins in this configuration, resulting in a dispersion characterized by two observed modes of degenerate spin waves. Furthermore, distinctive intersections between these modes occur at high-symmetry points M and K', suggesting the presence of magnetic Dirac nodes.

Our investigation continued into the impact of asymmetry, introduced through $\alpha'$ and $\beta'$ interactions, within the zigzag (ZZ) magnetic configuration. Higher asymmetry correlated with increased dispersion energy across the high symmetry pathway, influencing spin-wave propagation and retaining magnetic Dirac nodes at critical points like M and K'. Additionally, exploring the next nearest neighbor interaction ($\beta > 0$) revealed singular mode spin-waves without the potential for Dirac nodes. Stability heightened with rising $\beta$, resulting in elevated energy levels and distinct peaks at specific high-symmetry points, a consistent behavior even transitioning to $\beta < 0$. The double degeneracy of the next nearest neighbor interaction underlined its role in maintaining a spin-invariant nature, while asymmetry introduced through $\beta'$ resembled behaviors akin to anti-ferromagnetic configurations, devoid of Dirac modes. Our study illuminates how individual interactions and their asymmetries shape complex spin dynamics in the ZZ magnetic configuration, providing crucial insights for understanding and manipulating spin behaviors in analogous lattice structures.

Investigating the non-frustrated model highlighted how interactions seamlessly merged without competition, maintaining a harmonious 'content' configuration. Starting with the nearest neighbor interaction $\alpha$, we introduced the next nearest neighbor interaction $\beta$. As $\beta$ gradually approached equality with $\alpha$, observed changes unveiled an overall energy spectrum increase at peaks M and K', while troughs at $\Gamma$ became more defined. This emphasized the dominance of the nearest neighbor interaction dictating observed behaviors, with the next nearest neighbor enhancing the spin-wave strength and controlling crossover points, potentially due to a degeneracy effect observed in Fig.\ref{fig:Next Nearest Neighbor interaction}.
Continuing the exploration by introducing an AFM second nearest neighbor interaction $\alpha$' when $\alpha$=$\beta$, revealed a gradual increase in asymmetry and defined behavioral changes, evident from K-$\Gamma$ and M'-$\Gamma$, aligning with prior observations in Fig.\ref{fig:Nearest Neighbor interaction}. Further completion of the non-frustrated model with asymmetry among the next nearest neighbor interactions $\beta$' exhibited distinct deformations across the spin wave as energy increased to ~11 meV, maintaining stability without requiring easy-axis anisotropy.


Moving from the non-frustrated model, we observe a scenario where identical magnetic interactions led to competitive behavior, inducing frustration and resulting in a distinctive dispersion. Introducing AFM $\alpha$' disrupted symmetry, leading to an unstable dispersion until stabilized by the imposition of easy-axis anisotropy. This alteration caused a significant shift in the Dirac crossover point from ~6 meV to ~0.8 meV, showcasing a remarkable inversion in the behavior of the spin wave. The addition of further asymmetry, especially with the introduction of $\beta$', demonstrated shifts in energy up to ~8 meV, concurrently decreasing the required anisotropy for system stabilization. We then scrutinized variations in asymmetry, initiating next-nearest neighbor asymmetry $\beta$' and systematically introducing nearest neighbor asymmetry $\alpha$'. These meticulous analyses uncovered initial deformations in the spin wave, accompanied by a decrease in anisotropy as $\alpha$' increased. This process yielded refined peaks observed between M-K-$\Gamma$-K' and from M'-$\Gamma$, offering comprehensive insights into the intricate dynamics underlying frustration within interactions.

 Employing a frustrated model with asymmetric superexchange interactions and easy axis anisotropy, we analyzed inelastic neutron scattering data for $\alpha$-RuCl$_3$. While our model doesn't encompass Kitaev interactions completely, we approximated inelastic neutron scattering data \cite{Ran17PRL}, aiming to adjust model parameters for a better understanding of the Heisenberg interaction's role within this material. Within this framework, the Holstein-Primakoff expansion facilitates the characterization of spin excitations as magnons, allowing higher-order exchange parameters to encapsulate the competition inherent to the zig-zag configuration.

The current investigation into $\alpha$-RuCl$_3$ underlines its potential for fractionalized Kitaev physics and quantum spin liquid behavior. However, ongoing debates regarding its structural space group continue to impact our understanding of electronic and magnetic configurations. The magnetic behavior in $\alpha$-RuCl$_3$ stems from $Ru^{3+}$ ions' S=1/2 spins, exhibiting a zigzag pattern due to anti-parallel alignment along one axis and subsequent reversal along another. As the honeycomb lattice induces frustration in magnetic configurations, interactions become anisotropic and rely on spins relative to Ru-Ru bonds. Notably, our analysis reconfirms past experiments, emphasizing the prevalence of easy-axis anisotropy in monolayer samples\cite{Yang22NatMat}; moreover, our study strongly suggests that a frustrated Heisenberg interaction is a key determinant shaping the observed spin-wave behavior in $\alpha$-RuCl$_3$.

Additionally, our study highlights the notable observation that the breaking of inversion symmetry in $\alpha$-RuCl$_3$ is direction-dependent and produces distinctly visible Dirac cones. Our applied frustrated model for the zigzag magnetic configuration shows that this symmetry breaking occurs in two directions, aligning with our understanding of $\alpha$-RuCl$_3$'s electronic behavior. This paper contributes to comprehending magnetic interactions within the zig-zag honeycomb lattice, with direct implications for $\alpha$-RuCl$_3$. Our study imparts valuable insights into the lattice's spin dynamics, accentuating the pivotal roles played by diverse exchange interactions, asymmetry, and anisotropy in molding the system's behavior.

\bibliography{ZigZag-revised}

\begin{thebibliography}{53}%
\makeatletter
\providecommand \@ifxundefined [1]{%
 \@ifx{#1\undefined}
}%
\providecommand \@ifnum [1]{%
 \ifnum #1\expandafter \@firstoftwo
 \else \expandafter \@secondoftwo
 \fi
}%
\providecommand \@ifx [1]{%
 \ifx #1\expandafter \@firstoftwo
 \else \expandafter \@secondoftwo
 \fi
}%
\providecommand \natexlab [1]{#1}%
\providecommand \enquote  [1]{``#1''}%
\providecommand \bibnamefont  [1]{#1}%
\providecommand \bibfnamefont [1]{#1}%
\providecommand \citenamefont [1]{#1}%
\providecommand \href@noop [0]{\@secondoftwo}%
\providecommand \href [0]{\begingroup \@sanitize@url \@href}%
\providecommand \@href[1]{\@@startlink{#1}\@@href}%
\providecommand \@@href[1]{\endgroup#1\@@endlink}%
\providecommand \@sanitize@url [0]{\catcode `\\12\catcode `\$12\catcode
  `\&12\catcode `\#12\catcode `\^12\catcode `\_12\catcode `\%12\relax}%
\providecommand \@@startlink[1]{}%
\providecommand \@@endlink[0]{}%
\providecommand \url  [0]{\begingroup\@sanitize@url \@url }%
\providecommand \@url [1]{\endgroup\@href {#1}{\urlprefix }}%
\providecommand \urlprefix  [0]{URL }%
\providecommand \Eprint [0]{\href }%
\providecommand \doibase [0]{http://dx.doi.org/}%
\providecommand \selectlanguage [0]{\@gobble}%
\providecommand \bibinfo  [0]{\@secondoftwo}%
\providecommand \bibfield  [0]{\@secondoftwo}%
\providecommand \translation [1]{[#1]}%
\providecommand \BibitemOpen [0]{}%
\providecommand \bibitemStop [0]{}%
\providecommand \bibitemNoStop [0]{.\EOS\space}%
\providecommand \EOS [0]{\spacefactor3000\relax}%
\providecommand \BibitemShut  [1]{\csname bibitem#1\endcsname}%
\let\auto@bib@innerbib\@empty
\bibitem [{\citenamefont {Novoselov}\ \emph {et~al.}(2004)\citenamefont
  {Novoselov}, \citenamefont {Geim}, \citenamefont {Morozov}, \citenamefont
  {Jiang}, \citenamefont {Zhang}, \citenamefont {Dubonos}, \citenamefont
  {Grigorieva},\ and\ \citenamefont {Firsov}}]{Novoselov04AAAS}%
  \BibitemOpen
  \bibfield  {author} {\bibinfo {author} {\bibfnamefont {K.~S.}\ \bibnamefont
  {Novoselov}}, \bibinfo {author} {\bibfnamefont {A.~K.}\ \bibnamefont {Geim}},
  \bibinfo {author} {\bibfnamefont {S.~V.}\ \bibnamefont {Morozov}}, \bibinfo
  {author} {\bibfnamefont {D.}~\bibnamefont {Jiang}}, \bibinfo {author}
  {\bibfnamefont {Y.}~\bibnamefont {Zhang}}, \bibinfo {author} {\bibfnamefont
  {S.~V.}\ \bibnamefont {Dubonos}}, \bibinfo {author} {\bibfnamefont {I.~V.}\
  \bibnamefont {Grigorieva}}, \ and\ \bibinfo {author} {\bibfnamefont {A.~A.}\
  \bibnamefont {Firsov}},\ }\href@noop {} {\bibfield  {journal} {\bibinfo
  {journal} {Science}\ }\textbf {\bibinfo {volume} {306}},\ \bibinfo {pages}
  {666} (\bibinfo {year} {2004})}\BibitemShut {NoStop}%
\bibitem [{\citenamefont {Geim}\ and\ \citenamefont
  {Novoselov}(2007)}]{Geim07NatMat}%
  \BibitemOpen
  \bibfield  {author} {\bibinfo {author} {\bibfnamefont {A.}~\bibnamefont
  {Geim}}\ and\ \bibinfo {author} {\bibfnamefont {K.}~\bibnamefont
  {Novoselov}},\ }\href {\doibase 10.1038/nmat1849} {\bibfield  {journal}
  {\bibinfo  {journal} {Nature Materials}\ }\textbf {\bibinfo {volume} {6}},\
  \bibinfo {pages} {183} (\bibinfo {year} {2007})}\BibitemShut {NoStop}%
\bibitem [{\citenamefont {Banerjee}\ \emph
  {et~al.}(2016{\natexlab{a}})\citenamefont {Banerjee}, \citenamefont
  {Fransson}, \citenamefont {Black-Schaffer}, \citenamefont {\AA{}gren},\ and\
  \citenamefont {Balatsky}}]{Baner16PRB}%
  \BibitemOpen
  \bibfield  {author} {\bibinfo {author} {\bibfnamefont {S.}~\bibnamefont
  {Banerjee}}, \bibinfo {author} {\bibfnamefont {J.}~\bibnamefont {Fransson}},
  \bibinfo {author} {\bibfnamefont {A.~M.}\ \bibnamefont {Black-Schaffer}},
  \bibinfo {author} {\bibfnamefont {H.}~\bibnamefont {\AA{}gren}}, \ and\
  \bibinfo {author} {\bibfnamefont {A.~V.}\ \bibnamefont {Balatsky}},\
  }\href@noop {} {\bibfield  {journal} {\bibinfo  {journal} {Phys. Rev. B}\
  }\textbf {\bibinfo {volume} {93}},\ \bibinfo {pages} {134502} (\bibinfo
  {year} {2016}{\natexlab{a}})}\BibitemShut {NoStop}%
\bibitem [{\citenamefont {Chaloupka}\ \emph {et~al.}(2013)\citenamefont
  {Chaloupka}, \citenamefont {Jackeli},\ and\ \citenamefont
  {Khaliullin}}]{Chaloupka13PRL}%
  \BibitemOpen
  \bibfield  {author} {\bibinfo {author} {\bibfnamefont {J.~c.~v.}\
  \bibnamefont {Chaloupka}}, \bibinfo {author} {\bibfnamefont {G.}~\bibnamefont
  {Jackeli}}, \ and\ \bibinfo {author} {\bibfnamefont {G.}~\bibnamefont
  {Khaliullin}},\ }\href@noop {} {\bibfield  {journal} {\bibinfo  {journal}
  {Phys. Rev. Lett.}\ }\textbf {\bibinfo {volume} {110}},\ \bibinfo {pages}
  {097204} (\bibinfo {year} {2013})}\BibitemShut {NoStop}%
\bibitem [{\citenamefont {Chaloupka}\ and\ \citenamefont
  {Khaliullin}(2016)}]{Chaloupka16PRB}%
  \BibitemOpen
  \bibfield  {author} {\bibinfo {author} {\bibfnamefont {J.~c.~v.}\
  \bibnamefont {Chaloupka}}\ and\ \bibinfo {author} {\bibfnamefont
  {G.}~\bibnamefont {Khaliullin}},\ }\href@noop {} {\bibfield  {journal}
  {\bibinfo  {journal} {Phys. Rev. B}\ }\textbf {\bibinfo {volume} {94}},\
  \bibinfo {pages} {064435} (\bibinfo {year} {2016})}\BibitemShut {NoStop}%
\bibitem [{\citenamefont {Fransson}\ \emph {et~al.}(2016)\citenamefont
  {Fransson}, \citenamefont {Black-Schaffer},\ and\ \citenamefont
  {Balatsky}}]{Fransson16PRB}%
  \BibitemOpen
  \bibfield  {author} {\bibinfo {author} {\bibfnamefont {J.}~\bibnamefont
  {Fransson}}, \bibinfo {author} {\bibfnamefont {A.~M.}\ \bibnamefont
  {Black-Schaffer}}, \ and\ \bibinfo {author} {\bibfnamefont {A.~V.}\
  \bibnamefont {Balatsky}},\ }\href {\doibase 10.1103/PhysRevB.94.075401}
  {\bibfield  {journal} {\bibinfo  {journal} {Phys. Rev. B}\ }\textbf {\bibinfo
  {volume} {94}},\ \bibinfo {pages} {075401} (\bibinfo {year}
  {2016})}\BibitemShut {NoStop}%
\bibitem [{\citenamefont {Singh}\ and\ \citenamefont
  {Gegenwart}(2010)}]{Singh10PRB}%
  \BibitemOpen
  \bibfield  {author} {\bibinfo {author} {\bibfnamefont {Y.}~\bibnamefont
  {Singh}}\ and\ \bibinfo {author} {\bibfnamefont {P.}~\bibnamefont
  {Gegenwart}},\ }\href {\doibase 10.1103/PhysRevB.82.064412} {\bibfield
  {journal} {\bibinfo  {journal} {Phys. Rev. B}\ }\textbf {\bibinfo {volume}
  {82}},\ \bibinfo {pages} {064412} (\bibinfo {year} {2010})}\BibitemShut
  {NoStop}%
\bibitem [{\citenamefont {Boyko}\ \emph {et~al.}(2018)\citenamefont {Boyko},
  \citenamefont {Balatsky},\ and\ \citenamefont {Haraldsen}}]{Boyko18PRB}%
  \BibitemOpen
  \bibfield  {author} {\bibinfo {author} {\bibfnamefont {D.}~\bibnamefont
  {Boyko}}, \bibinfo {author} {\bibfnamefont {A.~V.}\ \bibnamefont {Balatsky}},
  \ and\ \bibinfo {author} {\bibfnamefont {J.~T.}\ \bibnamefont {Haraldsen}},\
  }\href@noop {} {\bibfield  {journal} {\bibinfo  {journal} {Phys. Rev. B}\
  }\textbf {\bibinfo {volume} {97}},\ \bibinfo {pages} {014433} (\bibinfo
  {year} {2018})}\BibitemShut {NoStop}%
\bibitem [{\citenamefont {Cayssol}(2013)}]{cayssol13CRP}%
  \BibitemOpen
  \bibfield  {author} {\bibinfo {author} {\bibfnamefont {J.}~\bibnamefont
  {Cayssol}},\ }\href@noop {} {\bibfield  {journal} {\bibinfo  {journal}
  {Comptes Rendus Physique}\ }\textbf {\bibinfo {volume} {14}},\ \bibinfo
  {pages} {760} (\bibinfo {year} {2013})},\ \bibinfo {note} {topological
  insulators / Isolants topologiques}\BibitemShut {NoStop}%
\bibitem [{\citenamefont {Gao}\ \emph {et~al.}(2021)\citenamefont {Gao},
  \citenamefont {Chen}, \citenamefont {Wang}, \citenamefont {Chen},
  \citenamefont {Zhong}, \citenamefont {Abernathy}, \citenamefont {Xiao},\ and\
  \citenamefont {Dai}}]{Gao21PRB}%
  \BibitemOpen
  \bibfield  {author} {\bibinfo {author} {\bibfnamefont {B.}~\bibnamefont
  {Gao}}, \bibinfo {author} {\bibfnamefont {T.}~\bibnamefont {Chen}}, \bibinfo
  {author} {\bibfnamefont {C.}~\bibnamefont {Wang}}, \bibinfo {author}
  {\bibfnamefont {L.}~\bibnamefont {Chen}}, \bibinfo {author} {\bibfnamefont
  {R.}~\bibnamefont {Zhong}}, \bibinfo {author} {\bibfnamefont {D.~L.}\
  \bibnamefont {Abernathy}}, \bibinfo {author} {\bibfnamefont {D.}~\bibnamefont
  {Xiao}}, \ and\ \bibinfo {author} {\bibfnamefont {P.}~\bibnamefont {Dai}},\
  }\href@noop {} {\bibfield  {journal} {\bibinfo  {journal} {Phys. Rev. B}\
  }\textbf {\bibinfo {volume} {104}},\ \bibinfo {pages} {214432} (\bibinfo
  {year} {2021})}\BibitemShut {NoStop}%
\bibitem [{\citenamefont {Cheng}\ \emph {et~al.}(2016)\citenamefont {Cheng},
  \citenamefont {Okamoto},\ and\ \citenamefont {Xiao}}]{Cheng16PRL}%
  \BibitemOpen
  \bibfield  {author} {\bibinfo {author} {\bibfnamefont {R.}~\bibnamefont
  {Cheng}}, \bibinfo {author} {\bibfnamefont {S.}~\bibnamefont {Okamoto}}, \
  and\ \bibinfo {author} {\bibfnamefont {D.}~\bibnamefont {Xiao}},\ }\href@noop
  {} {\bibfield  {journal} {\bibinfo  {journal} {Phys. Rev. Lett.}\ }\textbf
  {\bibinfo {volume} {117}},\ \bibinfo {pages} {217202} (\bibinfo {year}
  {2016})}\BibitemShut {NoStop}%
\bibitem [{\citenamefont {Haraldsen}\ and\ \citenamefont
  {Fishman}(2009)}]{Haraldsen09JoP}%
  \BibitemOpen
  \bibfield  {author} {\bibinfo {author} {\bibfnamefont {J.~T.}\ \bibnamefont
  {Haraldsen}}\ and\ \bibinfo {author} {\bibfnamefont {R.~S.}\ \bibnamefont
  {Fishman}},\ }\href@noop {} {\bibfield  {journal} {\bibinfo  {journal}
  {Journal of Physics: Condensed Matter}\ }\textbf {\bibinfo {volume} {21}},\
  \bibinfo {pages} {216001} (\bibinfo {year} {2009})}\BibitemShut {NoStop}%
\bibitem [{\citenamefont {Haraldsen}\ \emph {et~al.}(2009)\citenamefont
  {Haraldsen}, \citenamefont {Swanson}, \citenamefont {Alvarez},\ and\
  \citenamefont {Fishman}}]{Haraldsen09PRL}%
  \BibitemOpen
  \bibfield  {author} {\bibinfo {author} {\bibfnamefont {J.~T.}\ \bibnamefont
  {Haraldsen}}, \bibinfo {author} {\bibfnamefont {M.}~\bibnamefont {Swanson}},
  \bibinfo {author} {\bibfnamefont {G.}~\bibnamefont {Alvarez}}, \ and\
  \bibinfo {author} {\bibfnamefont {R.~S.}\ \bibnamefont {Fishman}},\ }\href
  {\doibase 10.1103/PhysRevLett.102.237204} {\bibfield  {journal} {\bibinfo
  {journal} {Phys. Rev. Lett.}\ }\textbf {\bibinfo {volume} {102}},\ \bibinfo
  {pages} {237204} (\bibinfo {year} {2009})}\BibitemShut {NoStop}%
\bibitem [{\citenamefont {Toth}\ and\ \citenamefont {Lake}(2015)}]{Toth15JoP}%
  \BibitemOpen
  \bibfield  {author} {\bibinfo {author} {\bibfnamefont {S.}~\bibnamefont
  {Toth}}\ and\ \bibinfo {author} {\bibfnamefont {B.}~\bibnamefont {Lake}},\
  }\href {\doibase 10.1088/0953-8984/27/16/166002} {\bibfield  {journal}
  {\bibinfo  {journal} {Journal of Physics: Condensed Matter}\ }\textbf
  {\bibinfo {volume} {27}},\ \bibinfo {pages} {166002} (\bibinfo {year}
  {2015})}\BibitemShut {NoStop}%
\bibitem [{\citenamefont {Auerbach}(2018)}]{AuerbachAPS2018}%
  \BibitemOpen
  \bibfield  {author} {\bibinfo {author} {\bibfnamefont {A.}~\bibnamefont
  {Auerbach}},\ }\href {\doibase 10.1103/PhysRevLett.121.066601} {\bibfield
  {journal} {\bibinfo  {journal} {Phys. Rev. Lett.}\ }\textbf {\bibinfo
  {volume} {121}},\ \bibinfo {pages} {066601} (\bibinfo {year}
  {2018})}\BibitemShut {NoStop}%
\bibitem [{\citenamefont {Cabra}\ \emph {et~al.}(2011)\citenamefont {Cabra},
  \citenamefont {Lamas},\ and\ \citenamefont {Rosales}}]{Cabra11PRB}%
  \BibitemOpen
  \bibfield  {author} {\bibinfo {author} {\bibfnamefont {D.~C.}\ \bibnamefont
  {Cabra}}, \bibinfo {author} {\bibfnamefont {C.~A.}\ \bibnamefont {Lamas}}, \
  and\ \bibinfo {author} {\bibfnamefont {H.~D.}\ \bibnamefont {Rosales}},\
  }\href@noop {} {\bibfield  {journal} {\bibinfo  {journal} {Phys. Rev. B}\
  }\textbf {\bibinfo {volume} {83}},\ \bibinfo {pages} {094506} (\bibinfo
  {year} {2011})}\BibitemShut {NoStop}%
\bibitem [{\citenamefont {Holstein}\ and\ \citenamefont
  {Primakoff}(1940)}]{Holstein40PR}%
  \BibitemOpen
  \bibfield  {author} {\bibinfo {author} {\bibfnamefont {T.}~\bibnamefont
  {Holstein}}\ and\ \bibinfo {author} {\bibfnamefont {H.}~\bibnamefont
  {Primakoff}},\ }\href {\doibase 10.1103/PhysRev.58.1098} {\bibfield
  {journal} {\bibinfo  {journal} {Phys. Rev.}\ }\textbf {\bibinfo {volume}
  {58}},\ \bibinfo {pages} {1098} (\bibinfo {year} {1940})}\BibitemShut
  {NoStop}%
\bibitem [{\citenamefont {Fouet}\ \emph {et~al.}(2001)\citenamefont {Fouet},
  \citenamefont {Sindzingre},\ and\ \citenamefont {Lhuillier}}]{Fouet01EPJB}%
  \BibitemOpen
  \bibfield  {author} {\bibinfo {author} {\bibfnamefont {J.}~\bibnamefont
  {Fouet}}, \bibinfo {author} {\bibfnamefont {P.}~\bibnamefont {Sindzingre}}, \
  and\ \bibinfo {author} {\bibfnamefont {C.}~\bibnamefont {Lhuillier}},\ }\href
  {\doibase 10.1007/s100510170273} {\bibfield  {journal} {\bibinfo  {journal}
  {The European Physical Journal B}\ }\textbf {\bibinfo {volume} {20}},\
  \bibinfo {pages} {241} (\bibinfo {year} {2001})}\BibitemShut {NoStop}%
\bibitem [{\citenamefont {Nieto}\ and\ \citenamefont
  {Truax}(1997)}]{Nieto97PoP}%
  \BibitemOpen
  \bibfield  {author} {\bibinfo {author} {\bibfnamefont {M.~M.}\ \bibnamefont
  {Nieto}}\ and\ \bibinfo {author} {\bibfnamefont {D.~R.}\ \bibnamefont
  {Truax}},\ }\href@noop {} {\bibfield  {journal} {\bibinfo  {journal}
  {Fortschritte der Physik/Progress of Physics}\ }\textbf {\bibinfo {volume}
  {45}},\ \bibinfo {pages} {145} (\bibinfo {year} {1997})}\BibitemShut
  {NoStop}%
\bibitem [{\citenamefont {Imamura}\ \emph {et~al.}(2004)\citenamefont
  {Imamura}, \citenamefont {Bruno},\ and\ \citenamefont
  {Utsumi}}]{Imamura04PRB}%
  \BibitemOpen
  \bibfield  {author} {\bibinfo {author} {\bibfnamefont {H.}~\bibnamefont
  {Imamura}}, \bibinfo {author} {\bibfnamefont {P.}~\bibnamefont {Bruno}}, \
  and\ \bibinfo {author} {\bibfnamefont {Y.}~\bibnamefont {Utsumi}},\
  }\href@noop {} {\bibfield  {journal} {\bibinfo  {journal} {Phys. Rev. B}\
  }\textbf {\bibinfo {volume} {69}},\ \bibinfo {pages} {121303} (\bibinfo
  {year} {2004})}\BibitemShut {NoStop}%
\bibitem [{\citenamefont {Trumper}\ \emph {et~al.}(2000)\citenamefont
  {Trumper}, \citenamefont {Capriotti},\ and\ \citenamefont
  {Sorella}}]{Trumper00PRB}%
  \BibitemOpen
  \bibfield  {author} {\bibinfo {author} {\bibfnamefont {A.~E.}\ \bibnamefont
  {Trumper}}, \bibinfo {author} {\bibfnamefont {L.}~\bibnamefont {Capriotti}},
  \ and\ \bibinfo {author} {\bibfnamefont {S.}~\bibnamefont {Sorella}},\ }\href
  {\doibase 10.1103/PhysRevB.61.11529} {\bibfield  {journal} {\bibinfo
  {journal} {Phys. Rev. B}\ }\textbf {\bibinfo {volume} {61}},\ \bibinfo
  {pages} {11529} (\bibinfo {year} {2000})}\BibitemShut {NoStop}%
\bibitem [{\citenamefont {Colpa}(1978)}]{Colpa78Physica}%
  \BibitemOpen
  \bibfield  {author} {\bibinfo {author} {\bibfnamefont {J.}~\bibnamefont
  {Colpa}},\ }\href@noop {} {\bibfield  {journal} {\bibinfo  {journal} {Physica
  A: Statistical Mechanics and its Applications}\ }\textbf {\bibinfo {volume}
  {93}},\ \bibinfo {pages} {327} (\bibinfo {year} {1978})}\BibitemShut
  {NoStop}%
\bibitem [{\citenamefont {Ozel}\ \emph {et~al.}(2019)\citenamefont {Ozel},
  \citenamefont {Belvin}, \citenamefont {Baldini}, \citenamefont {Kimchi},
  \citenamefont {Do}, \citenamefont {Choi},\ and\ \citenamefont
  {Gedik}}]{Ozel19PRB}%
  \BibitemOpen
  \bibfield  {author} {\bibinfo {author} {\bibfnamefont {I.~O.}\ \bibnamefont
  {Ozel}}, \bibinfo {author} {\bibfnamefont {C.~A.}\ \bibnamefont {Belvin}},
  \bibinfo {author} {\bibfnamefont {E.}~\bibnamefont {Baldini}}, \bibinfo
  {author} {\bibfnamefont {I.}~\bibnamefont {Kimchi}}, \bibinfo {author}
  {\bibfnamefont {S.}~\bibnamefont {Do}}, \bibinfo {author} {\bibfnamefont
  {K.-Y.}\ \bibnamefont {Choi}}, \ and\ \bibinfo {author} {\bibfnamefont
  {N.}~\bibnamefont {Gedik}},\ }\href@noop {} {\bibfield  {journal} {\bibinfo
  {journal} {Phys. Rev. B}\ }\textbf {\bibinfo {volume} {100}},\ \bibinfo
  {pages} {085108} (\bibinfo {year} {2019})}\BibitemShut {NoStop}%
\bibitem [{\citenamefont {Tierno}(2016)}]{Tierno16PRL}%
  \BibitemOpen
  \bibfield  {author} {\bibinfo {author} {\bibfnamefont {P.}~\bibnamefont
  {Tierno}},\ }\href {\doibase 10.1103/PhysRevLett.116.038303} {\bibfield
  {journal} {\bibinfo  {journal} {Phys. Rev. Lett.}\ }\textbf {\bibinfo
  {volume} {116}},\ \bibinfo {pages} {038303} (\bibinfo {year}
  {2016})}\BibitemShut {NoStop}%
\bibitem [{\citenamefont {Moessner}(2001)}]{Moessner01CJoP}%
  \BibitemOpen
  \bibfield  {author} {\bibinfo {author} {\bibfnamefont {R.}~\bibnamefont
  {Moessner}},\ }\href@noop {} {\bibfield  {journal} {\bibinfo  {journal}
  {Canadian Journal of Physics}\ }\textbf {\bibinfo {volume} {79}},\ \bibinfo
  {pages} {1283} (\bibinfo {year} {2001})}\BibitemShut {NoStop}%
\bibitem [{\citenamefont {Swanson}\ \emph {et~al.}(2009)\citenamefont
  {Swanson}, \citenamefont {Haraldsen},\ and\ \citenamefont
  {Fishman}}]{Swanson09PRB}%
  \BibitemOpen
  \bibfield  {author} {\bibinfo {author} {\bibfnamefont {M.}~\bibnamefont
  {Swanson}}, \bibinfo {author} {\bibfnamefont {J.~T.}\ \bibnamefont
  {Haraldsen}}, \ and\ \bibinfo {author} {\bibfnamefont {R.~S.}\ \bibnamefont
  {Fishman}},\ }\href@noop {} {\bibfield  {journal} {\bibinfo  {journal} {Phys.
  Rev. B}\ }\textbf {\bibinfo {volume} {79}},\ \bibinfo {pages} {184413}
  (\bibinfo {year} {2009})}\BibitemShut {NoStop}%
\bibitem [{\citenamefont {Beach}\ \emph {et~al.}(2009)\citenamefont {Beach},
  \citenamefont {Alet}, \citenamefont {Mambrini},\ and\ \citenamefont
  {Capponi}}]{Beach09PRB}%
  \BibitemOpen
  \bibfield  {author} {\bibinfo {author} {\bibfnamefont {K.~S.~D.}\
  \bibnamefont {Beach}}, \bibinfo {author} {\bibfnamefont {F.}~\bibnamefont
  {Alet}}, \bibinfo {author} {\bibfnamefont {M.}~\bibnamefont {Mambrini}}, \
  and\ \bibinfo {author} {\bibfnamefont {S.}~\bibnamefont {Capponi}},\
  }\href@noop {} {\bibfield  {journal} {\bibinfo  {journal} {Phys. Rev. B}\
  }\textbf {\bibinfo {volume} {80}},\ \bibinfo {pages} {184401} (\bibinfo
  {year} {2009})}\BibitemShut {NoStop}%
\bibitem [{\citenamefont {Ran}\ \emph {et~al.}(2017)\citenamefont {Ran},
  \citenamefont {Wang}, \citenamefont {Wang}, \citenamefont {Dong},
  \citenamefont {Ren}, \citenamefont {Bao}, \citenamefont {Li}, \citenamefont
  {Ma}, \citenamefont {Gan}, \citenamefont {Zhang}, \citenamefont {Park},
  \citenamefont {Deng}, \citenamefont {Danilkin}, \citenamefont {Yu},
  \citenamefont {Li},\ and\ \citenamefont {Wen}}]{Ran17PRL}%
  \BibitemOpen
  \bibfield  {author} {\bibinfo {author} {\bibfnamefont {K.}~\bibnamefont
  {Ran}}, \bibinfo {author} {\bibfnamefont {J.}~\bibnamefont {Wang}}, \bibinfo
  {author} {\bibfnamefont {W.}~\bibnamefont {Wang}}, \bibinfo {author}
  {\bibfnamefont {Z.-Y.}\ \bibnamefont {Dong}}, \bibinfo {author}
  {\bibfnamefont {X.}~\bibnamefont {Ren}}, \bibinfo {author} {\bibfnamefont
  {S.}~\bibnamefont {Bao}}, \bibinfo {author} {\bibfnamefont {S.}~\bibnamefont
  {Li}}, \bibinfo {author} {\bibfnamefont {Z.}~\bibnamefont {Ma}}, \bibinfo
  {author} {\bibfnamefont {Y.}~\bibnamefont {Gan}}, \bibinfo {author}
  {\bibfnamefont {Y.}~\bibnamefont {Zhang}}, \bibinfo {author} {\bibfnamefont
  {J.~T.}\ \bibnamefont {Park}}, \bibinfo {author} {\bibfnamefont
  {G.}~\bibnamefont {Deng}}, \bibinfo {author} {\bibfnamefont {S.}~\bibnamefont
  {Danilkin}}, \bibinfo {author} {\bibfnamefont {S.-L.}\ \bibnamefont {Yu}},
  \bibinfo {author} {\bibfnamefont {J.-X.}\ \bibnamefont {Li}}, \ and\ \bibinfo
  {author} {\bibfnamefont {J.}~\bibnamefont {Wen}},\ }\href@noop {} {\bibfield
  {journal} {\bibinfo  {journal} {Phys. Rev. Lett.}\ }\textbf {\bibinfo
  {volume} {118}},\ \bibinfo {pages} {107203} (\bibinfo {year}
  {2017})}\BibitemShut {NoStop}%
\bibitem [{\citenamefont {Johnson}\ \emph {et~al.}(2015)\citenamefont
  {Johnson}, \citenamefont {Williams}, \citenamefont {Haghighirad},
  \citenamefont {Singleton}, \citenamefont {Zapf}, \citenamefont {Manuel},
  \citenamefont {Mazin}, \citenamefont {Li}, \citenamefont {Jeschke},
  \citenamefont {Valent\'{\i}},\ and\ \citenamefont {Coldea}}]{Johnson15PRB}%
  \BibitemOpen
  \bibfield  {author} {\bibinfo {author} {\bibfnamefont {R.~D.}\ \bibnamefont
  {Johnson}}, \bibinfo {author} {\bibfnamefont {S.~C.}\ \bibnamefont
  {Williams}}, \bibinfo {author} {\bibfnamefont {A.~A.}\ \bibnamefont
  {Haghighirad}}, \bibinfo {author} {\bibfnamefont {J.}~\bibnamefont
  {Singleton}}, \bibinfo {author} {\bibfnamefont {V.}~\bibnamefont {Zapf}},
  \bibinfo {author} {\bibfnamefont {P.}~\bibnamefont {Manuel}}, \bibinfo
  {author} {\bibfnamefont {I.~I.}\ \bibnamefont {Mazin}}, \bibinfo {author}
  {\bibfnamefont {Y.}~\bibnamefont {Li}}, \bibinfo {author} {\bibfnamefont
  {H.~O.}\ \bibnamefont {Jeschke}}, \bibinfo {author} {\bibfnamefont
  {R.}~\bibnamefont {Valent\'{\i}}}, \ and\ \bibinfo {author} {\bibfnamefont
  {R.}~\bibnamefont {Coldea}},\ }\href@noop {} {\bibfield  {journal} {\bibinfo
  {journal} {Phys. Rev. B}\ }\textbf {\bibinfo {volume} {92}},\ \bibinfo
  {pages} {235119} (\bibinfo {year} {2015})}\BibitemShut {NoStop}%
\bibitem [{\citenamefont {Roslova}\ \emph {et~al.}(2019)\citenamefont
  {Roslova}, \citenamefont {Hunger}, \citenamefont {Bastien}, \citenamefont
  {Pohl}, \citenamefont {Haghighi}, \citenamefont {Wolter}, \citenamefont
  {Isaeva}, \citenamefont {Schwarz}, \citenamefont {Rellinghaus}, \citenamefont
  {Nielsch}, \citenamefont {Büchner},\ and\ \citenamefont
  {Doert}}]{Roslova19InoChem}%
  \BibitemOpen
  \bibfield  {author} {\bibinfo {author} {\bibfnamefont {M.}~\bibnamefont
  {Roslova}}, \bibinfo {author} {\bibfnamefont {J.}~\bibnamefont {Hunger}},
  \bibinfo {author} {\bibfnamefont {G.}~\bibnamefont {Bastien}}, \bibinfo
  {author} {\bibfnamefont {D.}~\bibnamefont {Pohl}}, \bibinfo {author}
  {\bibfnamefont {H.~M.}\ \bibnamefont {Haghighi}}, \bibinfo {author}
  {\bibfnamefont {A.~U.~B.}\ \bibnamefont {Wolter}}, \bibinfo {author}
  {\bibfnamefont {A.}~\bibnamefont {Isaeva}}, \bibinfo {author} {\bibfnamefont
  {U.}~\bibnamefont {Schwarz}}, \bibinfo {author} {\bibfnamefont
  {B.}~\bibnamefont {Rellinghaus}}, \bibinfo {author} {\bibfnamefont
  {K.}~\bibnamefont {Nielsch}}, \bibinfo {author} {\bibfnamefont
  {B.}~\bibnamefont {Büchner}}, \ and\ \bibinfo {author} {\bibfnamefont
  {T.}~\bibnamefont {Doert}},\ }\href@noop {} {\bibfield  {journal} {\bibinfo
  {journal} {Inorganic Chemistry}\ }\textbf {\bibinfo {volume} {58}},\ \bibinfo
  {pages} {6659} (\bibinfo {year} {2019})}\BibitemShut {NoStop}%
\bibitem [{\citenamefont {Banerjee}\ \emph
  {et~al.}(2016{\natexlab{b}})\citenamefont {Banerjee}, \citenamefont
  {Bridges}, \citenamefont {Yan}, \citenamefont {Aczel}, \citenamefont {Li},
  \citenamefont {Stone}, \citenamefont {Granroth}, \citenamefont {Lumsden},
  \citenamefont {Yiu}, \citenamefont {Knolle}, \citenamefont {Bhattacharjee},
  \citenamefont {Kovrizhin}, \citenamefont {Moessner}, \citenamefont {Tennant},
  \citenamefont {Mandrus},\ and\ \citenamefont {Nagler}}]{Baner16NatMat}%
  \BibitemOpen
  \bibfield  {author} {\bibinfo {author} {\bibfnamefont {A.}~\bibnamefont
  {Banerjee}}, \bibinfo {author} {\bibfnamefont {C.}~\bibnamefont {Bridges}},
  \bibinfo {author} {\bibfnamefont {J.-Q.}\ \bibnamefont {Yan}}, \bibinfo
  {author} {\bibfnamefont {A.}~\bibnamefont {Aczel}}, \bibinfo {author}
  {\bibfnamefont {L.}~\bibnamefont {Li}}, \bibinfo {author} {\bibfnamefont
  {M.}~\bibnamefont {Stone}}, \bibinfo {author} {\bibfnamefont
  {G.}~\bibnamefont {Granroth}}, \bibinfo {author} {\bibfnamefont
  {M.}~\bibnamefont {Lumsden}}, \bibinfo {author} {\bibfnamefont
  {Y.}~\bibnamefont {Yiu}}, \bibinfo {author} {\bibfnamefont {J.}~\bibnamefont
  {Knolle}}, \bibinfo {author} {\bibfnamefont {S.}~\bibnamefont
  {Bhattacharjee}}, \bibinfo {author} {\bibfnamefont {D.}~\bibnamefont
  {Kovrizhin}}, \bibinfo {author} {\bibfnamefont {R.}~\bibnamefont {Moessner}},
  \bibinfo {author} {\bibfnamefont {D.}~\bibnamefont {Tennant}}, \bibinfo
  {author} {\bibfnamefont {D.}~\bibnamefont {Mandrus}}, \ and\ \bibinfo
  {author} {\bibfnamefont {S.}~\bibnamefont {Nagler}},\ }\href {\doibase
  10.1038/nmat4604} {\bibfield  {journal} {\bibinfo  {journal} {Nature
  materials}\ }\textbf {\bibinfo {volume} {15}},\ \bibinfo {pages} {733—740}
  (\bibinfo {year} {2016}{\natexlab{b}})}\BibitemShut {NoStop}%
\bibitem [{\citenamefont {Savary}\ and\ \citenamefont
  {Balents}(2016)}]{Savary17RoPiP}%
  \BibitemOpen
  \bibfield  {author} {\bibinfo {author} {\bibfnamefont {L.}~\bibnamefont
  {Savary}}\ and\ \bibinfo {author} {\bibfnamefont {L.}~\bibnamefont
  {Balents}},\ }\href {\doibase 10.1088/0034-4885/80/1/016502} {\bibfield
  {journal} {\bibinfo  {journal} {Reports on Progress in Physics}\ }\textbf
  {\bibinfo {volume} {80}},\ \bibinfo {pages} {016502} (\bibinfo {year}
  {2016})}\BibitemShut {NoStop}%
\bibitem [{\citenamefont {Kim}\ \emph {et~al.}(2015)\citenamefont {Kim},
  \citenamefont {V.}, \citenamefont {Catuneanu},\ and\ \citenamefont
  {Kee}}]{Kim15PRB}%
  \BibitemOpen
  \bibfield  {author} {\bibinfo {author} {\bibfnamefont {H.-S.}\ \bibnamefont
  {Kim}}, \bibinfo {author} {\bibfnamefont {V.~S.}\ \bibnamefont {V.}},
  \bibinfo {author} {\bibfnamefont {A.}~\bibnamefont {Catuneanu}}, \ and\
  \bibinfo {author} {\bibfnamefont {H.-Y.}\ \bibnamefont {Kee}},\ }\href@noop
  {} {\bibfield  {journal} {\bibinfo  {journal} {Phys. Rev. B}\ }\textbf
  {\bibinfo {volume} {91}},\ \bibinfo {pages} {241110} (\bibinfo {year}
  {2015})}\BibitemShut {NoStop}%
\bibitem [{\citenamefont {Sandilands}\ \emph {et~al.}(2016)\citenamefont
  {Sandilands}, \citenamefont {Tian}, \citenamefont {Reijnders}, \citenamefont
  {Kim}, \citenamefont {Plumb}, \citenamefont {Kim}, \citenamefont {Kee},\ and\
  \citenamefont {Burch}}]{Sandilands16PRB}%
  \BibitemOpen
  \bibfield  {author} {\bibinfo {author} {\bibfnamefont {L.~J.}\ \bibnamefont
  {Sandilands}}, \bibinfo {author} {\bibfnamefont {Y.}~\bibnamefont {Tian}},
  \bibinfo {author} {\bibfnamefont {A.~A.}\ \bibnamefont {Reijnders}}, \bibinfo
  {author} {\bibfnamefont {H.-S.}\ \bibnamefont {Kim}}, \bibinfo {author}
  {\bibfnamefont {K.~W.}\ \bibnamefont {Plumb}}, \bibinfo {author}
  {\bibfnamefont {Y.-J.}\ \bibnamefont {Kim}}, \bibinfo {author} {\bibfnamefont
  {H.-Y.}\ \bibnamefont {Kee}}, \ and\ \bibinfo {author} {\bibfnamefont
  {K.~S.}\ \bibnamefont {Burch}},\ }\href@noop {} {\bibfield  {journal}
  {\bibinfo  {journal} {Phys. Rev. B}\ }\textbf {\bibinfo {volume} {93}},\
  \bibinfo {pages} {075144} (\bibinfo {year} {2016})}\BibitemShut {NoStop}%
\bibitem [{\citenamefont {Fletcher}\ \emph {et~al.}(1967)\citenamefont
  {Fletcher}, \citenamefont {Gardner}, \citenamefont {Fox},\ and\ \citenamefont
  {Topping}}]{Fletcher67JCSA}%
  \BibitemOpen
  \bibfield  {author} {\bibinfo {author} {\bibfnamefont {J.~M.}\ \bibnamefont
  {Fletcher}}, \bibinfo {author} {\bibfnamefont {W.~E.}\ \bibnamefont
  {Gardner}}, \bibinfo {author} {\bibfnamefont {A.~C.}\ \bibnamefont {Fox}}, \
  and\ \bibinfo {author} {\bibfnamefont {G.}~\bibnamefont {Topping}},\ }\href
  {\doibase 10.1039/J19670001038} {\bibfield  {journal} {\bibinfo  {journal}
  {J. Chem. Soc. A}\ ,\ \bibinfo {pages} {1038}} (\bibinfo {year}
  {1967})}\BibitemShut {NoStop}%
\bibitem [{\citenamefont {Park}\ \emph {et~al.}(2016)\citenamefont {Park},
  \citenamefont {Do}, \citenamefont {Choi}, \citenamefont {Jang}, \citenamefont
  {Jang}, \citenamefont {Schefer}, \citenamefont {Wu}, \citenamefont {Gardner},
  \citenamefont {Park}, \citenamefont {Park},\ and\ \citenamefont
  {Ji}}]{park2016emergence}%
  \BibitemOpen
  \bibfield  {author} {\bibinfo {author} {\bibfnamefont {S.~Y.}\ \bibnamefont
  {Park}}, \bibinfo {author} {\bibfnamefont {S.~H.}\ \bibnamefont {Do}},
  \bibinfo {author} {\bibfnamefont {K.~Y.}\ \bibnamefont {Choi}}, \bibinfo
  {author} {\bibfnamefont {D.}~\bibnamefont {Jang}}, \bibinfo {author}
  {\bibfnamefont {T.~H.}\ \bibnamefont {Jang}}, \bibinfo {author}
  {\bibfnamefont {J.}~\bibnamefont {Schefer}}, \bibinfo {author} {\bibfnamefont
  {C.~M.}\ \bibnamefont {Wu}}, \bibinfo {author} {\bibfnamefont {J.~S.}\
  \bibnamefont {Gardner}}, \bibinfo {author} {\bibfnamefont {J.~M.~S.}\
  \bibnamefont {Park}}, \bibinfo {author} {\bibfnamefont {J.~H.}\ \bibnamefont
  {Park}}, \ and\ \bibinfo {author} {\bibfnamefont {S.}~\bibnamefont {Ji}},\
  }\href@noop {} {\enquote {\bibinfo {title} {Emergence of the isotropic kitaev
  honeycomb lattice with two-dimensional ising universality in
  {$\alpha$}-rucl$_3$},}\ } (\bibinfo {year} {2016}),\ \Eprint
  {http://arxiv.org/abs/1609.05690} {arXiv:1609.05690 [cond-mat.mtrl-sci]}
  \BibitemShut {NoStop}%
\bibitem [{\citenamefont {Pai}\ \emph {et~al.}(2021)\citenamefont {Pai},
  \citenamefont {Marvinney}, \citenamefont {Feldman}, \citenamefont {Lerner},
  \citenamefont {Phang}, \citenamefont {Xiao}, \citenamefont {Yan},
  \citenamefont {Liang}, \citenamefont {Lapano}, \citenamefont {Brahlek},\ and\
  \citenamefont {Lawrie}}]{Pai2021JOPCC}%
  \BibitemOpen
  \bibfield  {author} {\bibinfo {author} {\bibfnamefont {Y.-Y.}\ \bibnamefont
  {Pai}}, \bibinfo {author} {\bibfnamefont {C.~E.}\ \bibnamefont {Marvinney}},
  \bibinfo {author} {\bibfnamefont {M.~A.}\ \bibnamefont {Feldman}}, \bibinfo
  {author} {\bibfnamefont {B.}~\bibnamefont {Lerner}}, \bibinfo {author}
  {\bibfnamefont {Y.~S.}\ \bibnamefont {Phang}}, \bibinfo {author}
  {\bibfnamefont {K.}~\bibnamefont {Xiao}}, \bibinfo {author} {\bibfnamefont
  {J.}~\bibnamefont {Yan}}, \bibinfo {author} {\bibfnamefont {L.}~\bibnamefont
  {Liang}}, \bibinfo {author} {\bibfnamefont {J.}~\bibnamefont {Lapano}},
  \bibinfo {author} {\bibfnamefont {M.}~\bibnamefont {Brahlek}}, \ and\
  \bibinfo {author} {\bibfnamefont {B.~J.}\ \bibnamefont {Lawrie}},\ }\href
  {\doibase 10.1021/acs.jpcc.1c07472} {\bibfield  {journal} {\bibinfo
  {journal} {The Journal of Physical Chemistry C}\ }\textbf {\bibinfo {volume}
  {125}},\ \bibinfo {pages} {25687–25694} (\bibinfo {year}
  {2021})}\BibitemShut {NoStop}%
\bibitem [{\citenamefont {Zhang}\ \emph
  {et~al.}(2023{\natexlab{a}})\citenamefont {Zhang}, \citenamefont {May},
  \citenamefont {Miao}, \citenamefont {Sales}, \citenamefont {Mandrus},
  \citenamefont {Nagler}, \citenamefont {McGuire},\ and\ \citenamefont
  {Yan}}]{HedaZhang2023PRM}%
  \BibitemOpen
  \bibfield  {author} {\bibinfo {author} {\bibfnamefont {H.}~\bibnamefont
  {Zhang}}, \bibinfo {author} {\bibfnamefont {A.~F.}\ \bibnamefont {May}},
  \bibinfo {author} {\bibfnamefont {H.}~\bibnamefont {Miao}}, \bibinfo {author}
  {\bibfnamefont {B.~C.}\ \bibnamefont {Sales}}, \bibinfo {author}
  {\bibfnamefont {D.~G.}\ \bibnamefont {Mandrus}}, \bibinfo {author}
  {\bibfnamefont {S.~E.}\ \bibnamefont {Nagler}}, \bibinfo {author}
  {\bibfnamefont {M.~A.}\ \bibnamefont {McGuire}}, \ and\ \bibinfo {author}
  {\bibfnamefont {J.}~\bibnamefont {Yan}},\ }\href {\doibase
  10.1103/PhysRevMaterials.7.114403} {\bibfield  {journal} {\bibinfo  {journal}
  {Phys. Rev. Mater.}\ }\textbf {\bibinfo {volume} {7}},\ \bibinfo {pages}
  {114403} (\bibinfo {year} {2023}{\natexlab{a}})}\BibitemShut {NoStop}%
\bibitem [{\citenamefont {Zhang}\ \emph
  {et~al.}(2023{\natexlab{b}})\citenamefont {Zhang}, \citenamefont
  {Tancogne-Dejean}, \citenamefont {Xian}, \citenamefont {Boström},
  \citenamefont {Claassen}, \citenamefont {Kennes},\ and\ \citenamefont
  {Rubio}}]{JinZhang2023NL}%
  \BibitemOpen
  \bibfield  {author} {\bibinfo {author} {\bibfnamefont {J.}~\bibnamefont
  {Zhang}}, \bibinfo {author} {\bibfnamefont {N.}~\bibnamefont
  {Tancogne-Dejean}}, \bibinfo {author} {\bibfnamefont {L.}~\bibnamefont
  {Xian}}, \bibinfo {author} {\bibfnamefont {E.~V.}\ \bibnamefont {Boström}},
  \bibinfo {author} {\bibfnamefont {M.}~\bibnamefont {Claassen}}, \bibinfo
  {author} {\bibfnamefont {D.~M.}\ \bibnamefont {Kennes}}, \ and\ \bibinfo
  {author} {\bibfnamefont {A.}~\bibnamefont {Rubio}},\ }\href {\doibase
  10.1021/acs.nanolett.3c02668} {\bibfield  {journal} {\bibinfo  {journal}
  {Nano Letters}\ }\textbf {\bibinfo {volume} {23}},\ \bibinfo {pages} {8712}
  (\bibinfo {year} {2023}{\natexlab{b}})},\ \bibinfo {note} {pMID:
  37695730}\BibitemShut {NoStop}%
\bibitem [{\citenamefont {Bera}\ \emph {et~al.}(2017)\citenamefont {Bera},
  \citenamefont {Yusuf}, \citenamefont {Kumar},\ and\ \citenamefont
  {Ritter}}]{Bera17PRB}%
  \BibitemOpen
  \bibfield  {author} {\bibinfo {author} {\bibfnamefont {A.~K.}\ \bibnamefont
  {Bera}}, \bibinfo {author} {\bibfnamefont {S.~M.}\ \bibnamefont {Yusuf}},
  \bibinfo {author} {\bibfnamefont {A.}~\bibnamefont {Kumar}}, \ and\ \bibinfo
  {author} {\bibfnamefont {C.}~\bibnamefont {Ritter}},\ }\href@noop {}
  {\bibfield  {journal} {\bibinfo  {journal} {Phys. Rev. B}\ }\textbf {\bibinfo
  {volume} {95}},\ \bibinfo {pages} {094424} (\bibinfo {year}
  {2017})}\BibitemShut {NoStop}%
\bibitem [{\citenamefont {Loidl}\ \emph {et~al.}(2021)\citenamefont {Loidl},
  \citenamefont {Lunkenheimer},\ and\ \citenamefont {Tsurkan}}]{Loidl21JoP}%
  \BibitemOpen
  \bibfield  {author} {\bibinfo {author} {\bibfnamefont {A.}~\bibnamefont
  {Loidl}}, \bibinfo {author} {\bibfnamefont {P.}~\bibnamefont {Lunkenheimer}},
  \ and\ \bibinfo {author} {\bibfnamefont {V.}~\bibnamefont {Tsurkan}},\ }\href
  {\doibase 10.1088/1361-648X/ac1bcf} {\bibfield  {journal} {\bibinfo
  {journal} {Journal of Physics: Condensed Matter}\ }\textbf {\bibinfo {volume}
  {33}},\ \bibinfo {pages} {443004} (\bibinfo {year} {2021})}\BibitemShut
  {NoStop}%
\bibitem [{\citenamefont {Warzanowski}\ \emph {et~al.}(2020)\citenamefont
  {Warzanowski}, \citenamefont {Borgwardt}, \citenamefont {Hopfer},
  \citenamefont {Attig}, \citenamefont {Koethe}, \citenamefont {Becker},
  \citenamefont {Tsurkan}, \citenamefont {Loidl}, \citenamefont {Hermanns},
  \citenamefont {van Loosdrecht},\ and\ \citenamefont
  {Gr\"uninger}}]{War20PRR}%
  \BibitemOpen
  \bibfield  {author} {\bibinfo {author} {\bibfnamefont {P.}~\bibnamefont
  {Warzanowski}}, \bibinfo {author} {\bibfnamefont {N.}~\bibnamefont
  {Borgwardt}}, \bibinfo {author} {\bibfnamefont {K.}~\bibnamefont {Hopfer}},
  \bibinfo {author} {\bibfnamefont {J.}~\bibnamefont {Attig}}, \bibinfo
  {author} {\bibfnamefont {T.~C.}\ \bibnamefont {Koethe}}, \bibinfo {author}
  {\bibfnamefont {P.}~\bibnamefont {Becker}}, \bibinfo {author} {\bibfnamefont
  {V.}~\bibnamefont {Tsurkan}}, \bibinfo {author} {\bibfnamefont
  {A.}~\bibnamefont {Loidl}}, \bibinfo {author} {\bibfnamefont
  {M.}~\bibnamefont {Hermanns}}, \bibinfo {author} {\bibfnamefont {P.~H.~M.}\
  \bibnamefont {van Loosdrecht}}, \ and\ \bibinfo {author} {\bibfnamefont
  {M.}~\bibnamefont {Gr\"uninger}},\ }\href@noop {} {\bibfield  {journal}
  {\bibinfo  {journal} {Phys. Rev. Res.}\ }\textbf {\bibinfo {volume} {2}},\
  \bibinfo {pages} {042007} (\bibinfo {year} {2020})}\BibitemShut {NoStop}%
\bibitem [{\citenamefont {Winter}\ \emph
  {et~al.}(2017{\natexlab{a}})\citenamefont {Winter}, \citenamefont {Tsirlin},
  \citenamefont {Daghofer}, \citenamefont {van~den Brink}, \citenamefont
  {Singh}, \citenamefont {Gegenwart},\ and\ \citenamefont {Valent{\'{\i}
  }}}]{Winter17JoP}%
  \BibitemOpen
  \bibfield  {author} {\bibinfo {author} {\bibfnamefont {S.~M.}\ \bibnamefont
  {Winter}}, \bibinfo {author} {\bibfnamefont {A.~A.}\ \bibnamefont {Tsirlin}},
  \bibinfo {author} {\bibfnamefont {M.}~\bibnamefont {Daghofer}}, \bibinfo
  {author} {\bibfnamefont {J.}~\bibnamefont {van~den Brink}}, \bibinfo {author}
  {\bibfnamefont {Y.}~\bibnamefont {Singh}}, \bibinfo {author} {\bibfnamefont
  {P.}~\bibnamefont {Gegenwart}}, \ and\ \bibinfo {author} {\bibfnamefont
  {R.}~\bibnamefont {Valent{\'{\i} }}},\ }\href@noop {} {\bibfield  {journal}
  {\bibinfo  {journal} {Journal of Physics: Condensed Matter}\ }\textbf
  {\bibinfo {volume} {29}},\ \bibinfo {pages} {493002} (\bibinfo {year}
  {2017}{\natexlab{a}})}\BibitemShut {NoStop}%
\bibitem [{\citenamefont {Winter}\ \emph
  {et~al.}(2017{\natexlab{b}})\citenamefont {Winter}, \citenamefont {Riedl},
  \citenamefont {Maksimov}, \citenamefont {Chernyshev}, \citenamefont
  {Honecker},\ and\ \citenamefont {Valent{\'{\i}}}}]{Winter17NatComm}%
  \BibitemOpen
  \bibfield  {author} {\bibinfo {author} {\bibfnamefont {S.~M.}\ \bibnamefont
  {Winter}}, \bibinfo {author} {\bibfnamefont {K.}~\bibnamefont {Riedl}},
  \bibinfo {author} {\bibfnamefont {P.~A.}\ \bibnamefont {Maksimov}}, \bibinfo
  {author} {\bibfnamefont {A.~L.}\ \bibnamefont {Chernyshev}}, \bibinfo
  {author} {\bibfnamefont {A.}~\bibnamefont {Honecker}}, \ and\ \bibinfo
  {author} {\bibfnamefont {R.}~\bibnamefont {Valent{\'{\i}}}},\ }\href@noop {}
  {\bibfield  {journal} {\bibinfo  {journal} {Nature Communications}\ }\textbf
  {\bibinfo {volume} {8}} (\bibinfo {year} {2017}{\natexlab{b}})}\BibitemShut
  {NoStop}%
\bibitem [{\citenamefont {Kubota}\ \emph {et~al.}(2015)\citenamefont {Kubota},
  \citenamefont {Tanaka}, \citenamefont {Ono}, \citenamefont {Narumi},\ and\
  \citenamefont {Kindo}}]{Kubota15PRB}%
  \BibitemOpen
  \bibfield  {author} {\bibinfo {author} {\bibfnamefont {Y.}~\bibnamefont
  {Kubota}}, \bibinfo {author} {\bibfnamefont {H.}~\bibnamefont {Tanaka}},
  \bibinfo {author} {\bibfnamefont {T.}~\bibnamefont {Ono}}, \bibinfo {author}
  {\bibfnamefont {Y.}~\bibnamefont {Narumi}}, \ and\ \bibinfo {author}
  {\bibfnamefont {K.}~\bibnamefont {Kindo}},\ }\href@noop {} {\bibfield
  {journal} {\bibinfo  {journal} {Phys. Rev. B}\ }\textbf {\bibinfo {volume}
  {91}},\ \bibinfo {pages} {094422} (\bibinfo {year} {2015})}\BibitemShut
  {NoStop}%
\bibitem [{\citenamefont {Bachus}\ \emph {et~al.}(2020)\citenamefont {Bachus},
  \citenamefont {Kaib}, \citenamefont {Tokiwa}, \citenamefont {Jesche},
  \citenamefont {Tsurkan}, \citenamefont {Loidl}, \citenamefont {Winter},
  \citenamefont {Tsirlin}, \citenamefont {Valent\'{\i}},\ and\ \citenamefont
  {Gegenwart}}]{Bachus20PRL}%
  \BibitemOpen
  \bibfield  {author} {\bibinfo {author} {\bibfnamefont {S.}~\bibnamefont
  {Bachus}}, \bibinfo {author} {\bibfnamefont {D.~A.~S.}\ \bibnamefont {Kaib}},
  \bibinfo {author} {\bibfnamefont {Y.}~\bibnamefont {Tokiwa}}, \bibinfo
  {author} {\bibfnamefont {A.}~\bibnamefont {Jesche}}, \bibinfo {author}
  {\bibfnamefont {V.}~\bibnamefont {Tsurkan}}, \bibinfo {author} {\bibfnamefont
  {A.}~\bibnamefont {Loidl}}, \bibinfo {author} {\bibfnamefont {S.~M.}\
  \bibnamefont {Winter}}, \bibinfo {author} {\bibfnamefont {A.~A.}\
  \bibnamefont {Tsirlin}}, \bibinfo {author} {\bibfnamefont {R.}~\bibnamefont
  {Valent\'{\i}}}, \ and\ \bibinfo {author} {\bibfnamefont {P.}~\bibnamefont
  {Gegenwart}},\ }\href@noop {} {\bibfield  {journal} {\bibinfo  {journal}
  {Phys. Rev. Lett.}\ }\textbf {\bibinfo {volume} {125}},\ \bibinfo {pages}
  {097203} (\bibinfo {year} {2020})}\BibitemShut {NoStop}%
\bibitem [{\citenamefont {Sears}\ \emph {et~al.}(2015)\citenamefont {Sears},
  \citenamefont {Songvilay}, \citenamefont {Plumb}, \citenamefont {Clancy},
  \citenamefont {Qiu}, \citenamefont {Zhao}, \citenamefont {Parshall},\ and\
  \citenamefont {Kim}}]{Sears15PRB}%
  \BibitemOpen
  \bibfield  {author} {\bibinfo {author} {\bibfnamefont {J.~A.}\ \bibnamefont
  {Sears}}, \bibinfo {author} {\bibfnamefont {M.}~\bibnamefont {Songvilay}},
  \bibinfo {author} {\bibfnamefont {K.~W.}\ \bibnamefont {Plumb}}, \bibinfo
  {author} {\bibfnamefont {J.~P.}\ \bibnamefont {Clancy}}, \bibinfo {author}
  {\bibfnamefont {Y.}~\bibnamefont {Qiu}}, \bibinfo {author} {\bibfnamefont
  {Y.}~\bibnamefont {Zhao}}, \bibinfo {author} {\bibfnamefont {D.}~\bibnamefont
  {Parshall}}, \ and\ \bibinfo {author} {\bibfnamefont {Y.-J.}\ \bibnamefont
  {Kim}},\ }\href@noop {} {\bibfield  {journal} {\bibinfo  {journal} {Phys.
  Rev. B}\ }\textbf {\bibinfo {volume} {91}},\ \bibinfo {pages} {144420}
  (\bibinfo {year} {2015})}\BibitemShut {NoStop}%
\bibitem [{\citenamefont {Luo}\ and\ \citenamefont {Kee}(2022)}]{Luo22PRB}%
  \BibitemOpen
  \bibfield  {author} {\bibinfo {author} {\bibfnamefont {Q.}~\bibnamefont
  {Luo}}\ and\ \bibinfo {author} {\bibfnamefont {H.-Y.}\ \bibnamefont {Kee}},\
  }\href@noop {} {\bibfield  {journal} {\bibinfo  {journal} {Phys. Rev. B}\
  }\textbf {\bibinfo {volume} {105}},\ \bibinfo {pages} {174435} (\bibinfo
  {year} {2022})}\BibitemShut {NoStop}%
\bibitem [{\citenamefont {Plumb}\ \emph {et~al.}(2014)\citenamefont {Plumb},
  \citenamefont {Clancy}, \citenamefont {Sandilands}, \citenamefont {Shankar},
  \citenamefont {Hu}, \citenamefont {Burch}, \citenamefont {Kee},\ and\
  \citenamefont {Kim}}]{Plumb14PRB}%
  \BibitemOpen
  \bibfield  {author} {\bibinfo {author} {\bibfnamefont {K.~W.}\ \bibnamefont
  {Plumb}}, \bibinfo {author} {\bibfnamefont {J.~P.}\ \bibnamefont {Clancy}},
  \bibinfo {author} {\bibfnamefont {L.~J.}\ \bibnamefont {Sandilands}},
  \bibinfo {author} {\bibfnamefont {V.~V.}\ \bibnamefont {Shankar}}, \bibinfo
  {author} {\bibfnamefont {Y.~F.}\ \bibnamefont {Hu}}, \bibinfo {author}
  {\bibfnamefont {K.~S.}\ \bibnamefont {Burch}}, \bibinfo {author}
  {\bibfnamefont {H.-Y.}\ \bibnamefont {Kee}}, \ and\ \bibinfo {author}
  {\bibfnamefont {Y.-J.}\ \bibnamefont {Kim}},\ }\href@noop {} {\bibfield
  {journal} {\bibinfo  {journal} {Phys. Rev. B}\ }\textbf {\bibinfo {volume}
  {90}},\ \bibinfo {pages} {041112} (\bibinfo {year} {2014})}\BibitemShut
  {NoStop}%
\bibitem [{\citenamefont {Yang}\ \emph {et~al.}(2022)\citenamefont {Yang},
  \citenamefont {Goh}, \citenamefont {Sung}, \citenamefont {Ye}, \citenamefont
  {Biswas}, \citenamefont {Kaib}, \citenamefont {Dhakal}, \citenamefont {Yan},
  \citenamefont {Li}, \citenamefont {Jiang}, \citenamefont {Chen},
  \citenamefont {Lei}, \citenamefont {He}, \citenamefont {Valent{\'{\i}}},
  \citenamefont {Winter}, \citenamefont {Hovden},\ and\ \citenamefont
  {Tsen}}]{Yang22NatMat}%
  \BibitemOpen
  \bibfield  {author} {\bibinfo {author} {\bibfnamefont {B.}~\bibnamefont
  {Yang}}, \bibinfo {author} {\bibfnamefont {Y.~M.}\ \bibnamefont {Goh}},
  \bibinfo {author} {\bibfnamefont {S.~H.}\ \bibnamefont {Sung}}, \bibinfo
  {author} {\bibfnamefont {G.}~\bibnamefont {Ye}}, \bibinfo {author}
  {\bibfnamefont {S.}~\bibnamefont {Biswas}}, \bibinfo {author} {\bibfnamefont
  {D.~A.~S.}\ \bibnamefont {Kaib}}, \bibinfo {author} {\bibfnamefont
  {R.}~\bibnamefont {Dhakal}}, \bibinfo {author} {\bibfnamefont
  {S.}~\bibnamefont {Yan}}, \bibinfo {author} {\bibfnamefont {C.}~\bibnamefont
  {Li}}, \bibinfo {author} {\bibfnamefont {S.}~\bibnamefont {Jiang}}, \bibinfo
  {author} {\bibfnamefont {F.}~\bibnamefont {Chen}}, \bibinfo {author}
  {\bibfnamefont {H.}~\bibnamefont {Lei}}, \bibinfo {author} {\bibfnamefont
  {R.}~\bibnamefont {He}}, \bibinfo {author} {\bibfnamefont {R.}~\bibnamefont
  {Valent{\'{\i}}}}, \bibinfo {author} {\bibfnamefont {S.~M.}\ \bibnamefont
  {Winter}}, \bibinfo {author} {\bibfnamefont {R.}~\bibnamefont {Hovden}}, \
  and\ \bibinfo {author} {\bibfnamefont {A.~W.}\ \bibnamefont {Tsen}},\
  }\href@noop {} {\bibfield  {journal} {\bibinfo  {journal} {Nature Materials}\
  }\textbf {\bibinfo {volume} {22}},\ \bibinfo {pages} {50} (\bibinfo {year}
  {2022})}\BibitemShut {NoStop}%
\bibitem [{\citenamefont {Wang}\ \emph {et~al.}(2022)\citenamefont {Wang},
  \citenamefont {Liu}, \citenamefont {Zheng}, \citenamefont {Zhao},
  \citenamefont {Yang}, \citenamefont {Wang}, \citenamefont {Yang},
  \citenamefont {Wu},\ and\ \citenamefont {Gao}}]{Wang22DirectOO}%
  \BibitemOpen
  \bibfield  {author} {\bibinfo {author} {\bibfnamefont {Z.}~\bibnamefont
  {Wang}}, \bibinfo {author} {\bibfnamefont {L.}~\bibnamefont {Liu}}, \bibinfo
  {author} {\bibfnamefont {H.}~\bibnamefont {Zheng}}, \bibinfo {author}
  {\bibfnamefont {M.-Q.}\ \bibnamefont {Zhao}}, \bibinfo {author}
  {\bibfnamefont {K.}~\bibnamefont {Yang}}, \bibinfo {author} {\bibfnamefont
  {C.}~\bibnamefont {Wang}}, \bibinfo {author} {\bibfnamefont {F.}~\bibnamefont
  {Yang}}, \bibinfo {author} {\bibfnamefont {H.}~\bibnamefont {Wu}}, \ and\
  \bibinfo {author} {\bibfnamefont {C.}~\bibnamefont {Gao}},\ }\href
  {https://api.semanticscholar.org/CorpusID:263513819} {\bibfield  {journal}
  {\bibinfo  {journal} {Nanoscale}\ } (\bibinfo {year} {2022})}\BibitemShut
  {NoStop}%
\bibitem [{\citenamefont {Goerbig}\ and\ \citenamefont
  {Montambaux}(2014)}]{Goerbig14dirac}%
  \BibitemOpen
  \bibfield  {author} {\bibinfo {author} {\bibfnamefont {M.~O.}\ \bibnamefont
  {Goerbig}}\ and\ \bibinfo {author} {\bibfnamefont {G.}~\bibnamefont
  {Montambaux}},\ }\href@noop {} {\bibfield  {journal} {\bibinfo  {journal}
  {Progress in Mathematical Physics}\ } (\bibinfo {year} {2014})},\ \Eprint
  {http://arxiv.org/abs/1410.4098} {1410.4098} \BibitemShut {NoStop}%
\bibitem [{\citenamefont {Kim}\ \emph {et~al.}(2020)\citenamefont {Kim},
  \citenamefont {Jeong}, \citenamefont {Park}, \citenamefont {Masuda},
  \citenamefont {Asai}, \citenamefont {Itoh}, \citenamefont {Kim},
  \citenamefont {Wildes},\ and\ \citenamefont {Park}}]{Kim20PRB}%
  \BibitemOpen
  \bibfield  {author} {\bibinfo {author} {\bibfnamefont {C.}~\bibnamefont
  {Kim}}, \bibinfo {author} {\bibfnamefont {J.}~\bibnamefont {Jeong}}, \bibinfo
  {author} {\bibfnamefont {P.}~\bibnamefont {Park}}, \bibinfo {author}
  {\bibfnamefont {T.}~\bibnamefont {Masuda}}, \bibinfo {author} {\bibfnamefont
  {S.}~\bibnamefont {Asai}}, \bibinfo {author} {\bibfnamefont {S.}~\bibnamefont
  {Itoh}}, \bibinfo {author} {\bibfnamefont {H.-S.}\ \bibnamefont {Kim}},
  \bibinfo {author} {\bibfnamefont {A.}~\bibnamefont {Wildes}}, \ and\ \bibinfo
  {author} {\bibfnamefont {J.-G.}\ \bibnamefont {Park}},\ }\href@noop {}
  {\bibfield  {journal} {\bibinfo  {journal} {Phys. Rev. B}\ }\textbf {\bibinfo
  {volume} {102}},\ \bibinfo {pages} {184429} (\bibinfo {year}
  {2020})}\BibitemShut {NoStop}%
\end{thebibliography}%

\end{document}